\documentclass[final,3p,onecolumn]{elsarticle}

\usepackage{amssymb}
\usepackage{wrapfig}
\usepackage{threeparttable}
\usepackage{subcaption}
\usepackage[]{hyperref}
\usepackage{placeins}
\usepackage{floatrow}
\usepackage{booktabs}
\usepackage{graphicx}
\usepackage{epstopdf}
\usepackage{MnSymbol}
\usepackage{wasysym}
\usepackage{ifsym}
\usepackage{color}
\usepackage{prettyref}
\usepackage{wrapfig}
 
  \newrefformat{eq}{\hyperref[{#1}]{Eq.~\textup{(\ref{#1})}}}
  \newrefformat{cha}{\hyperref[{#1}]{Chapter~\ref{#1}}}  
  \newrefformat{sec}{\hyperref[{#1}]{Section~\ref{#1}}}
  \newrefformat{app}{\hyperref[{#1}]{Appendix~\ref{#1}}}
  \newrefformat{tab}{\hyperref[{#1}]{Table~\ref{#1}}}
  \newrefformat{fig}{\hyperref[{#1}]{Fig.~\ref{#1}}}

\journal{Acta Astronautica-Under Review}

\begin{document}
\sloppy

\begin{frontmatter}

\title{Asteroid mining with small spacecraft and its economic feasibility}

\author[label2]{Pablo Calla\corref{cor1}}
\address[label2]{International Space University, 1 Rue Jean-Dominique Cassini, 67400 Illkirch-Graffenstaden, France}
\ead{pablo.calla@community.isunet.edu}

\author[label1]{Dan Fries}
\address[label1]{Initiative for Interstellar Studies, Bone Mill, New Street,  Charfield, GL12 8ES, United Kingdom}
\ead{dfries@gatech.edu}

\author[label2]{Chris Welch}
\ead{chris.welch@isunet.edu}

\begin{abstract}
Asteroid mining offers the possibility to revolutionize supply of resources vital for human civilization. Preliminary analysis suggests that Near-Earth Asteroids (NEA) contain enough volatile and high value minerals to make the mining process economically feasible. Considering possible applications, specifically the mining of water in space has become a major focus for near-term options. Most proposed projects for asteroid mining involve spacecraft based on traditional designs resulting in large, monolithic and expensive systems.

An alternative approach is presented in this paper, basing the asteroid mining process on multiple small spacecraft. To the best knowledge of the authors, only limited analysis of the asteroid mining capability of small spacecraft has been conducted. This paper explores the possibility to perform asteroid mining operations with spacecraft that have a mass under 500 kg and deliver 100 kg of water per trip. The mining process considers water extraction through microwave heating with an efficiency of 2 Wh/g.The proposed, small spacecraft can reach NEAs within a range of $\sim 0.03$ AU relative to earth's orbit, offering a delta V of 437 m/s per one-way trip.

A high-level systems engineering and economic analysis provides a closed spacecraft design as a baseline and puts the cost of the proposed spacecraft at \$ 113.6 million/unit. The results indicate that more than one hundred spacecraft and their successful operation for over five years are required to achieve a financial break-even point. Pros and cons of using small spacecraft swarms are highlighted and the uncertainties associated with cost and profit of space related business ventures are analyzed. 
\end{abstract}


\end{frontmatter}

\section{Introduction}

Asteroids are celestial bodies that are of fundamental scientific importance for uncovering the formation, composition and evolution of the solar system \cite{badescu2013asteroids}. Moreover, mining an asteroid for useful resources is a concept that even predates modern space programs, as an idea initially proposed in the early $20^{th}$ century by Konstantin Tsiolkovsky. More recent analysis suggests that specifically Near-Earth Asteroids (NEAs) are close enough and could contain trillions of dollars worth of precious metals and minerals, potentially making the endeavor feasible \cite{hellgren2016asteroid,sanchez2011asteroid,ASTRA2010}.Useful reservoirs of important substances may be found, such as water, metals and semiconductors \cite{sanchez2012assessment}. 

The extraction of volatiles is currently the most realistic near-term asteroid mining application. Therefore, several concepts for extraction and supply of water were developed recently \cite{badescu2013asteroids}. These concepts consider water extraction for refueling of spacecraft, radiation shielding, and potable water for life support systems in outer space \cite{hellgren2016asteroid}. 

In the last twenty years, a vast amount of data and results from space missions have been collected. Observations from spacecraft are mainly used to complement theories and findings which were deduced from ground based asteroid data \cite{badescu2013asteroids}. Although a full scale exploitation of space resources has not been achieved yet, some minimal asteroid samples have been retrieved for analysis and testing on earth. The number of discovered NEAs goes beyond 15000, with an average of 30 new asteroids discovered per week \cite{JPL-CNEOS}. However, estimates from ground based observations do not guarantee the accurate composition of asteroid candidates. Therefore, spacecraft are required for in-situ measurements complementing the data and establishing a clear candidate for exploitation. Current missions for asteroid mining consider spacecraft prospection as a first step before the extraction process.

Prospection itself usually falls into three different phases \cite{badescu2013asteroids}: discovery, remote characterization, local characterization. These last two characterization phases are endeavors currently pursued by asteroid mining companies using small spacecraft\cite{PlanetRes}. However, recent advances in the miniaturization of spacecraft components and mining equipment may allow for a more cost effective and reliable approach to mine NEAs overall. 

We propose and analyze a mission architecture focusing on the utilization of small spacecraft for the asteroid mining cycle. This includes local prospecting, mining, and return of relevant substances. Focusing on the extraction of volatiles as a first step, we conduct a survey of water mining techniques, other relevant technologies, and past missions. Using standard space systems engineering techniques a trade-off analysis is conducted to select a suitable mission architecture and spacecraft design. We identify the high level challenges facing asteroid mining, highlight where technological improvements are required, and present a road map for implementation. Conducting an analysis of the costs involved in establishing mining operations, constraints are derived on the economic feasibility of asteroid mining using the proposed architecture. The constraints reveal that the number of spacecraft and the target market for the retrieved volatiles is essential to achieve a break-even point.

\section{Heritage Mission Architecture}

Space missions to asteroids provide an accurate description of their composition, but more prospection missions are required to determine individual candidates for mineral exploitation. Besides exploration, mining techniques have also been developed but not yet tested in space. There are several mining approaches that vary in function of the asteroid size, as for small asteroids, whole asteroid capturing is a more feasible option, rather than for larger objects, in which extracting chunks of material is more reliable \cite{zacny2013asteroid}.

\subsection{Current Technology: Sample Retrieval}
 
There is significant interest in sample retrieval missions, having in mind that their characterization not only provides a deeper insight into the Solar System, but also represents a technological challenge for space exploration. Moreover, several space macro-engineering missions proposed are designed mainly to capture part or an entire asteroid and return to a useful orbit for its processing \cite{badescu2013asteroids}. 

The Keck Institute of Space Studies \cite{studies2012asteroid} mentions that there are five categories of benefits from the return of an asteroid sample or a full asteroid retrieval, which are: (1)  synergy with near-term human exploration, (2) expansion of international cooperation, (3) synergy with planetary defense, (4) exploitation of asteroid resources, (5) public engagement.

\begin{figure*}[t!]
	\centering
	\begin{subfigure}[t]{0.45\textwidth}
		\centering
		\includegraphics[width=\textwidth]{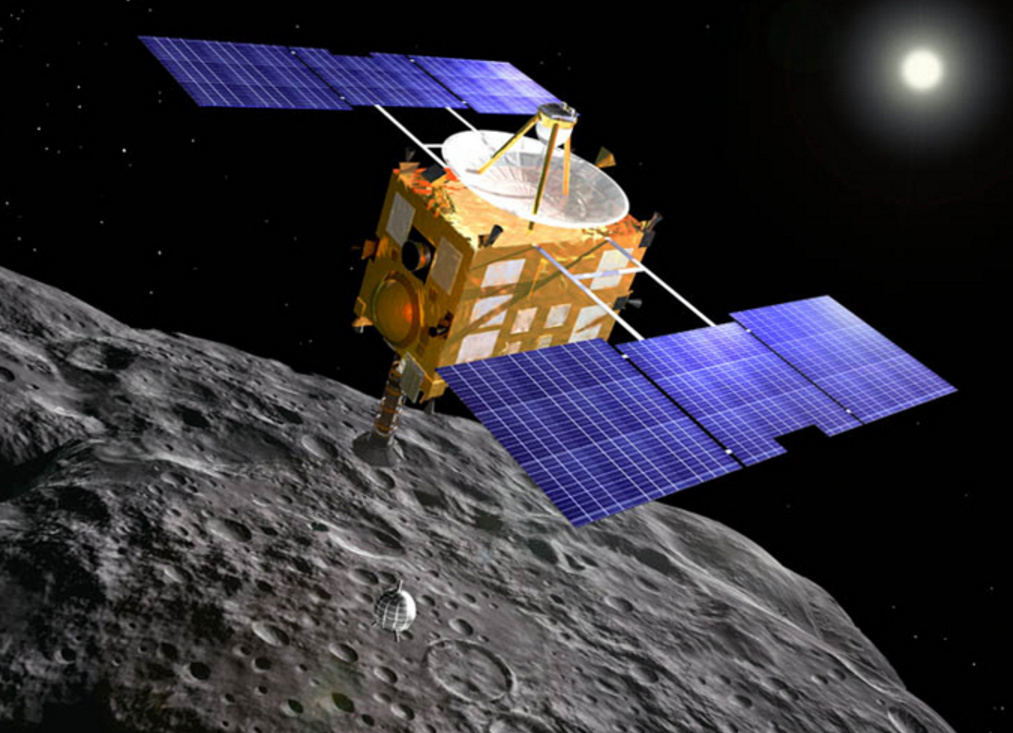}
		\caption{Artist's impression of Hayabusa 1. NASA Planetary Science Division, NASA JPL}
		\label{fig:Hayabusa_1}
	\end{subfigure}%
	~ 
	\begin{subfigure}[t]{0.46\textwidth}
		\centering
		 \includegraphics[width=\textwidth]{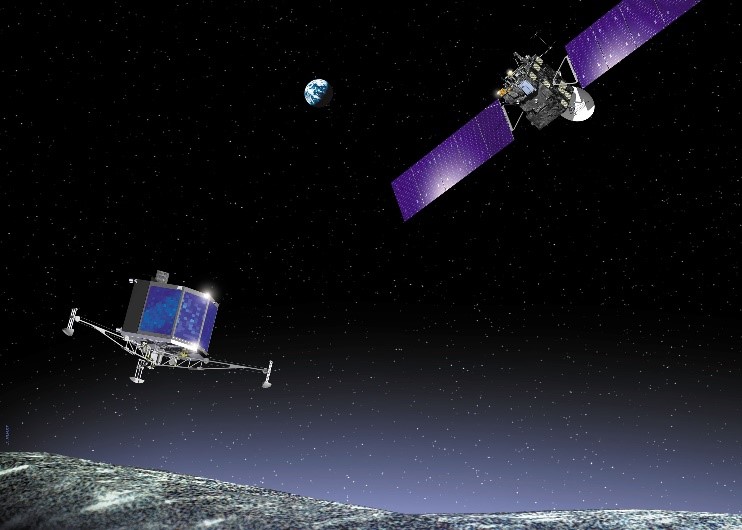}
		\caption{Artist's impression of the Rosetta mission. Space in images. ESA, 2015}
		\label{fig:Rosetta}
	\end{subfigure}
	\caption{Two examples of past missions to asteroids that included sample return, remote sensing, and/or landing on the asteroid.}
\end{figure*}

The past and present missions reviewed are Hayabusa 1 and 2 (see \prettyref{fig:Hayabusa_1}), Stardust, and Osiris Rex.

Sample return missions from asteroids were successful in the past as proven by the Hayabusa and the Stardust missions. It is possible to return a very small amount of material from outer space in order to be studied, however no approach has been made to try to test a technique in which large quantities of material can be extracted. Therefore, the establishment of asteroid mining requires the testing of mining techniques in space. Both techniques considered in past missions use collectors that cannot be adapted for larger scale extraction and processing purposes. Whats more, contemporary missions focus heavily on the science aspect of exploring asteroids. For example, Osiris-Rex techniques for sample retrieval are not scalable and depend on the amount of nitrogen the mission demands, limiting its functionality. This concept is not feasible for actual asteroid mining. Nevertheless, rendezvous and soft-landing techniques have been proven successfully. 

\subsection{Current Technology: Remote Sensing}

''The goal is to find ore, not merely a concentration of some minable resource"  \cite{badescu2013asteroids}. Ore refers to material that is commercially profitable and it can be precious metals, helium-3, water, organics or others. Prospecting is the first step to establish if the material abundance in an asteroid is potentially profitable, this is done by analyzing the data coming from observed asteroids, which provide information that can be interpreted to gain knowledge on the spin rate, size, shape, albedo reflection and to determine the type of asteroid. The techniques of prospecting or remote sensing are the same technology as the remote sensing satellites used on Earth \cite{hellgren2016asteroid}.

The following past and present missions were surveyed: Deep Space 1, NEAR Shoemaker, Rosetta (see \prettyref{fig:Rosetta}), WISE, and Dawn.

There have been several missions related to studying the characteristics of asteroids in the past years and also in the present. Many of the characteristics have been identified by ground observations and the missions complemented the data by close observation of some bodies. Although many asteroids have been identified, there is little detailed information about their composition. The missions sent had primary targets either in deep space or in the asteroid belt. Near Earth asteroids have not been analyzed deeply by any of these past missions. In order to pursue a space mining venture, the information about the bodies that are closer (NEAs) is critical to identify potential candidates for certain resource exploitation.

\subsection{Future Missions}
It is also important to review planned future mission designs to understand challenges, technology options and the purpose of these missions better. The following future missions were surveyed: Asteroid redirect mission (NASA, canceled), Prospector 1 (DSI), Arkyd Prospectors (Planetary Resources), Hedgehog (NASA), Robotic Asteroid Prospector, Asteroid Provided In-Situ Supplies.

The future mission concepts by two companies (Deep Space Industries and Planetary Resources) involved  in asteroid mining so far propose prospecting spacecraft only. Other approaches for asteroid mining involve capturing mechanisms aiming at very small asteroids of 20 m or less. These rely on relatively big spacecraft with the capability to catch and de-orbit the asteroid, returning it to a lunar or Earth orbit. There are a few small spacecraft concepts proposed for asteroid mining, but these consider a mother (bigger) spacecraft that carries them \cite{zacny2013asteroid}.

Although the feasibility analysis in several papers state that asteroid mining could be economically feasible, the technologies are not mature \cite{badescu2013asteroids}. Especially the process of anchoring and extraction is still a challenge.


\subsection{Water Mining Techniques}

The concept of water extraction involves finding water reservoirs in NEA’s, extract the water, process, and transport it to a location of value where a depot or processing plant is available. Water exists in the form of hydrated minerals and sometimes as ice, all of which can be refined into fuel, water for life support systems, air and radiation shielding. According to Dula \& Zhang \cite{dula2015space} the main customers that would consider water to be valuable in space will include all explorers preferring to buy less expensive fuel in space instead of transferring it from earth.

\subsubsection{Asteroid Water}

The first step is to identify the water reservoirs in asteroids. As an analogy of what it is expected to be found in asteroids, the celestial bodies’ composition is compared to the results found on Earth in terms of the elements they may contain and their water holding capacity.

\begin{wrapfigure}[26]{r}{0.5\textwidth}
	\includegraphics[width=\textwidth]{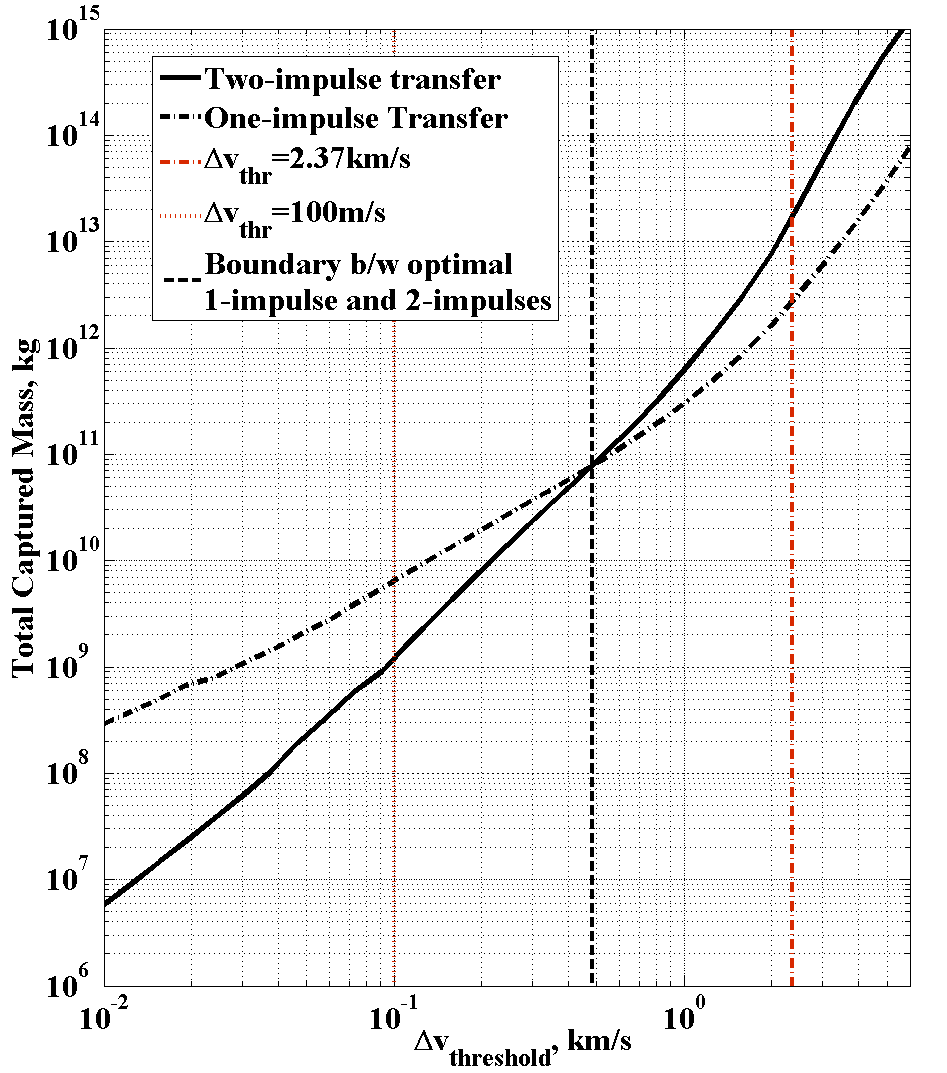}
	\caption{Expected water resource pool in near-earth C-type asteroids. (Sanchez \& McInnes (2011) \cite{sanchez2011asteroid}, figure reproduced with permission of the authors.)}
	\label{fig:water_pool}
\end{wrapfigure}

If hydrated carbonaceous asteroids in the Near Earth Asteroids population are considered as the only water carrying objects (around 10\% population) and assuming that these objects carry around 8.5\% water, the water in near Earth space can be estimated \cite{ross2001}. The estimates by Sanchez \& McInnes \cite{sanchez2011asteroid} are shown in \prettyref{fig:water_pool}.

The numbers are based on study results of asteroid composition determined by observing C-class asteroids for many years through an absorption band in their reflectance spectra, near 3 $\mu m$. The importance of asteroid prospectors to have an accurate estimate is highlighted before initiating an extraction process. Nevertheless, numerical estimates of water availability are subject to some uncertainty and might need adjustment in the future. An example for a slightly different range of numerical estimates has been published recently by A.S. Rivkin and F. E. DeMeo \cite{rivkin2019many}. A comparison of relevant numerical estimates shows that usage of their numbers would not change the conclusions presented in this paper significantly.

\subsubsection{Techniques Applicable in Space}

Unfortunately, no mining techniques have been developed specifically for a zero-/micro-gravity environment, yet. Nonetheless, different concepts have been studied and some solutions proposed.

Extracting water in barren planets such as Mars has also gained interest as a defining factor for human settlements. Given the fact that a few missions have landed on Mars, some concepts for water extraction were pro-posed. These concepts can also be applied to other celestial bodies such as NEA asteroids.

According to Wiens et al. \cite{wiens2001water} , potential designs for water extraction by heating include:
\begin{itemize}
  \item \textbf{Inclined Pipes:} Electrical heating elements heat the soil in a rotating inclined pipe. The released vapor would rise from the soil, travel the inside surface of the pipe and exit on top. The dehydrated soil would pass out the bot-tom of the pipe.
  \item \textbf{Kettles / Pots:} Soil is placed in an electrical heater inside a kettle releasing vapor. Then, vapor is condensed and collected as liquid water.
  \item \textbf{Sifters:} Soil passing through an electrically heated sifting screen releases vapor.
  \item \textbf{Funnels:} A funnel and a conveyor belt are used to heat the soil and release vapor.
  \item \textbf{Focused Light:} Focused sunlight to release water vapor from a portion of the soil.
  \item \textbf{Microwaves:} Heat the soil bound water via high power radio waves. The microwaves apply energy to the water directly and don't require heating the soil unlike conventional methods.
\end{itemize}

For any of these techniques, a cold trap needs to be attached in order to condense the vapor and collect liquid water. 

After assessing past mission architectures and taking into account the top level requirements for the asteroid mining venture with small spacecraft, the following architecture is proposed and displayed in Figure~\ref{fig:proposed_architecture2}:

After launch, the single or multiple spacecraft will be set in a parking orbit before entering a trajectory for rendezvous. The trajectory may consider a direct cruising to the asteroid. Once it arrives to the asteroid, it must determine the asteroid characteristics and adapt to that environment before the landing attempt. Maneuvers for landing are performed and an anchoring technique secures the spacecraft to the surface. Once fixed, the extraction is performed. The mineral extracted (water) will be processed in-situ or only stored, according to later designs. Following the operations, detachment is performed and stabilization for the transportation to a cis-lunar or Earth orbit facility.

\subsubsection{Selection}

A trade-off analysis is performed to select the water mining technique most suitable for the goals of this study, i.e. the usage of small spacecraft. A quantitative analysis is achieved by using a decision matrix. The water extraction techniques considered are based on existing studies by Bernold, Wiens et al., the Keck Institute for Space Studies, and Sercel \cite{bernold2013,wiens2001water,studies2012asteroid,sercel2016}. They are shown in \prettyref{tab:tech-compare} with their respective advantages and disadvantages.

\begin{table}[h!]
	\centering
	\resizebox{\textwidth}{!}{%
		\begin{tabular}{|l|l|l|}
			\hline
			\multicolumn{1}{|c}{\textbf{Water extraction technique}}    
			     & \multicolumn{1}{|c|}{\textbf{Advantages}}                                                                                          & \multicolumn{1}{c|}{\textbf{Disadvantages}}                                                                                                                            \\ \hline
			\textbf{Vacuum drying(Pneumatic System)} & \begin{tabular}[c]{@{}l@{}}Very Reliable \\ Few moving parts\end{tabular}                                  & \begin{tabular}[c]{@{}l@{}}Need to carry compressed air and other instruments \\ Hard to implement \\ Hardly scalable \\ Complex\end{tabular} \\ \hline
			\textbf{Hot air or steam drying}         & \begin{tabular}[c]{@{}l@{}}Very reliable \\ Few moving parts \\ Simple\end{tabular}                        & \begin{tabular}[c]{@{}l@{}}Need to carry compressed air and other instruments \\ Hard to implement\end{tabular}                                 \\ \hline
			\textbf{Solar drying (focused light)}    & \begin{tabular}[c]{@{}l@{}}Very Reliable\\ High temperature in small area\\ Feasible\end{tabular}          & \begin{tabular}[c]{@{}l@{}}Sunlight dependent \\ No night operation \\ Alignment required\end{tabular}                                          \\ \hline
			\textbf{Inclined pipes heating}          & \begin{tabular}[c]{@{}l@{}}Few moving parts \\ Continuous operation \\ Easy to collect vapor\end{tabular} & \begin{tabular}[c]{@{}l@{}}High energy required\\ High mass\end{tabular}                                                                                   \\ \hline
			\textbf{Kettle/Pot heating}              & \begin{tabular}[c]{@{}l@{}}Simple\\ Few moving parts\\ Amount vs time\end{tabular}                        & \begin{tabular}[c]{@{}l@{}}Insulation requirement \\ High energy required \\ High mass\end{tabular}                                             \\ \hline
			\textbf{Sifter heating}                  & \begin{tabular}[c]{@{}l@{}}Simple\\ No moving parts\end{tabular}                                           & \begin{tabular}[c]{@{}l@{}}Gravity Dependent \\ High clogging risk \\ High energy\\ Fast heating, less efficient\end{tabular}                  \\ \hline
			\textbf{Funnel heating}                  & \begin{tabular}[c]{@{}l@{}}Simple\\ No moving parts\\ Small amount of soil required\end{tabular}           & \begin{tabular}[c]{@{}l@{}}Gravity Dependent \\ Clogging risk \\ High energy\end{tabular}                                                         \\ \hline
			\textbf{Conveyor Belt (drum drying)}     & \begin{tabular}[c]{@{}l@{}}Continuous Operation \\ Very efficient\\ Very reliable\end{tabular}            & \begin{tabular}[c]{@{}l@{}}Gravity Dependent \\ Several moving parts \\ High energy \\ High mass\end{tabular}                                   \\ \hline
			\textbf{Microwave drying}                & \begin{tabular}[c]{@{}l@{}}Compact\\ Good efficiency \\ Few moving parts Reliable\end{tabular}            & \begin{tabular}[c]{@{}l@{}}Relatively high energy requirement\\ Relatively high mass\end{tabular}                                                    \\ \hline
			\textbf{Capturing and heating}           & \begin{tabular}[c]{@{}l@{}}Very efficient \\ Continuous Operation\end{tabular}                             & \begin{tabular}[c]{@{}l@{}}Sunlight dependent \\ Difficult to scale \\ Alignment required \\ Very hard to implement\end{tabular}               \\ \hline
	\end{tabular}}
	\caption{Extracting water technologies comparison.}
	\label{tab:tech-compare}
\end{table}

The decision criteria will consider a scale from 1 to 10 for each parameter and a weighting factor based on the importance of each parameter was set. The parameters considered to be evaluated are:

\begin{itemize}
	\item Scalability. The water technique needs to be scalable as it needs to be suitable for a small spacecraft. Higher score if it can be scalable. Weight = 1.5
	\item Complexity. The method’s complexity adds a relative risk to the mission. The higher the complexity, the higher the risk of failure. A higher score means that the method is simple. Weight = 1.2
	\item Technological Feasibility. Evaluation whether a technology is currently feasible to work in the asteroid environ-ment. High score if it was proven in similar conditions. Weight = 1.5
	\item Durability. The technique needs to be robust enough to work properly and repeatedly in a non- ideal scenario that could be encountered in an asteroid. High score if it is most likely durable. Weight = 1.
\end{itemize}

\prettyref{tab:decision-mat} in \hyperref[app:AppendixA]{App. A} contains the values assigned to each technique and the total score obtained. In the presented study, the highest score is obtained by the microwave drying technique. Although the focused light technique for heating also obtained a good score, it would not be entirely applicable for small spacecraft as it does not seem scalable enough. Microwaves are scalable and feasible to work in a harsh environment besides the technology being relatively simple and durable. Therefore, this method was selected in the present project. Some advantages of using the microwave techniques are also described by Wiens et al. \cite{wiens2001water}: greater efficiency than thermal energy sources, directly heat the bound water while not wasting energy to heat the soil, few moving parts, reliable, and no warm up period.

\section{Mission Architecture}

\begin{figure}[htb]
 \includegraphics[width=1\textwidth]{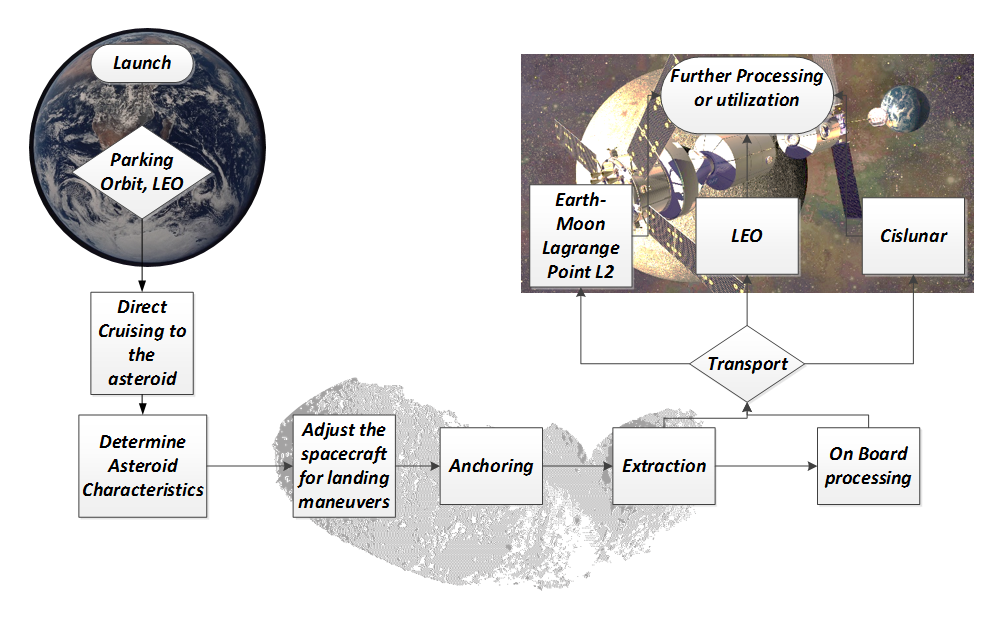}
 \caption{Proposed architecture for asteroid mining using small spacecraft.}
 \label{fig:proposed_architecture2}
\end{figure}

After assessing past system architectures (Section 2) we propose the mission flowchart presented in \prettyref{fig:proposed_architecture2} for asteroid mining utilizing small spacecraft in a generic approach in order to outline the step required. This mission architecture does assume that previous prospecting missions have been sent to a sufficient number of candidate asteroids to ensure that a large enough amount of water is present for extraction, and that other asteroid properties (such as angular velocity) are within the range of the mining spacecraft operating parameters.

According to the proposed mission architecture, either a single or multiple spacecraft will be positioned in a parking orbit before entering a trajectory for asteroid rendezvous. Upon its arrival at the asteroid, it will verify asteroid characteristics such as spin rate, surface mineral composition and overall shape.  It will then use this information to identify potential mining sites to land at.  The information shall be sent to Earth and the landing spots confirmed by ground analysis, considering that the spacecraft will not be able to perform image analysis and autonomous identification. Spacecraft adaptation, which refers to reorientation with respect to the asteroid characteristics, is then performed in order to continue with a soft landing. The landing maneuvers culminate with the spacecraft anchoring itself to the asteroid. The landing process is envisioned to occur in several hours (soft landing), thus, even a communication delay of several minutes will not represent a significant problem.

Once the spacecraft is secured, the initial drilling may begin, followed by positioning the microwave equipment, and finally the water extraction itself.  The operations will not be continuous, as the spacecraft may lose power during each eclipse, depending on the landing location.  In that scenario, the payload power will be cut once the solar panels are producing insufficient power to continue operations, and then restarted when the solar panels regain power after the eclipse period.  

Once extraction is complete and the water stored, the spacecraft will detach from the asteroid and reorient itself for the transit phase. Once it reaches the target Earth orbit, it will be tele-operated, if necessary, to position it for docking with a processing facility.  Once the cargo is delivered, the spacecraft will be inspected for any maintenance issues, and then repeat this cycle either at the original asteroid, or a new target.

\subsection{Mission and Spacecraft Requirements}\label{sec:MissionCraftReq}

After analyzing past asteroid mining concepts and defining the most appropriate water extraction technique, we need to make sure the system can fulfill the mission while avoiding over-design. To  this end, the following top-level requirements define in broad terms the functions and operations that the small spacecraft should perform \cite{larson1992space}.

\textbf{T1. The spacecraft should have a small structure:} The spacecraft configuration should correspond to the parameters that define a small spacecraft for this application, i.e. a mass of $<500$ kg according to NASA Ames documents \cite{agasid2015}.

\textbf{T2. The spacecraft should be dimensioned in order to obtain around 100 kg of water:} Up until now, there is no estimate of the actual demand of water in space. This makes it hard to determine the necessary quantity returned per asteroid run for a given operational cost. A value that has been utilized for calculations in past projects is 100 kg \cite{zacny2013reaching}, this quantity should be obtainable if the asteroid is least 3 meters in diameter ~\cite{zacny2013asteroid}.

\textbf{T3. The spacecraft should be able to be refueled in-situ with the extracted water and be able to operate with a water based propulsion system:} The spacecraft should be able to resupply itself to ensure steady operation and to mitigate possible losses due to lack of propellant. Therefore, it should carry a water based propulsion system or a water compatible system.

\textbf{T4. The spacecraft should carry imaging instruments in order to determine the best possible landing site:} The best possible landing site is determined by the spinning rate, the surface mineral content and the shape of the asteroid.

\textbf{T5. The spacecraft should incorporate reusable anchoring, storage and extraction systems to support mining operations.}

\textbf{T6. The spacecraft should have a high degree of autonomy to facilitate operations.}

\textbf{T7. The maximum travel distance of the spacecraft should be less than 0.1 AU from Earth:} Near Earth Objects (NEO’s) are asteroids and comets with perihelion distance of less than 1.3 AU. However, there are a large number of close approaches that are accessible for rendezvous at 0.1 AU, which limits the required propulsive capabilities.

\textbf{T8. The spacecraft shall be capable of delivering the collected water to a space station upon returning to an appropriate orbit:} Once the material has been extracted, it has to be delivered to the appropriate orbit. There, either docking or rendezvous might be required to deliver the material to a spacecraft or space station. 

\textbf{T9. The spacecraft cost must be minimized to improve the likelihood of economic feasibility.}

\textbf{T10. The spacecraft shall contain a backup propulsion system in order to be able to return to Earth if the water extraction is unsuccessful.}

\textbf{T11. The spacecraft shall be able to repeat mining trips several times before being decommissioned.}

\subsection{Mission Duration}

\begin{wrapfigure}{r}{0.6\textwidth}
	\includegraphics[width=\textwidth]{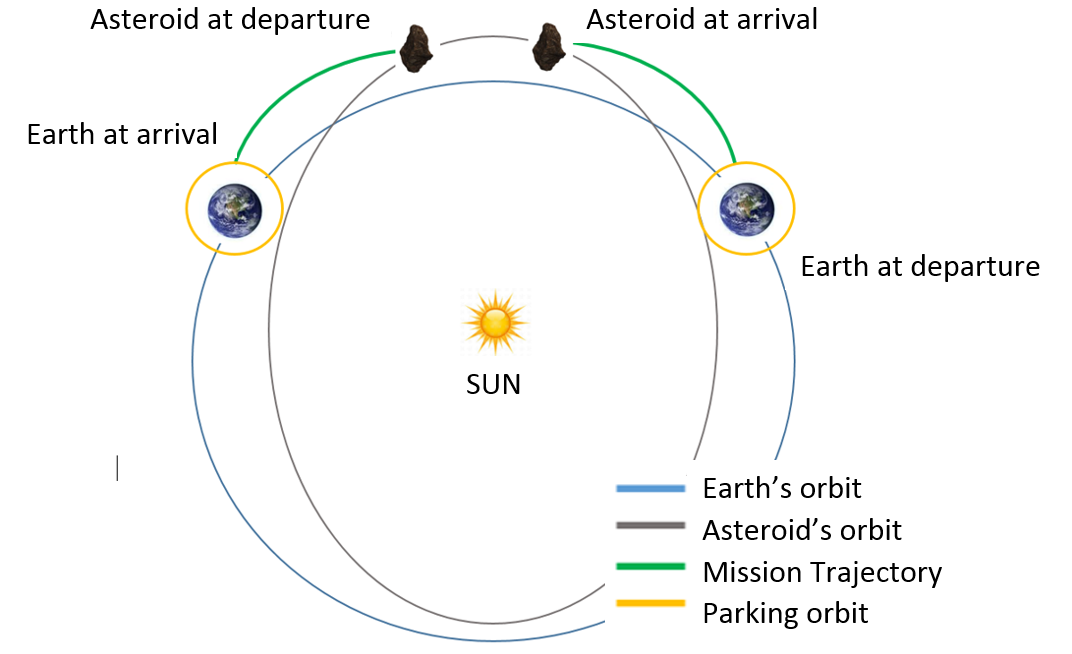}
	\caption{Proposed mission trajectory}
	\label{fig:proposed_trajectory}
\end{wrapfigure}

Past missions for rendezvous and sample collection have used the concept of ``rapid return missions" \cite{barbee2010comprehensive}, where the entire mission duration is between 6 and 9 months. The same concept can be applied for a low cost mining operation. However to ensure sufficient water is mined the total mission duration is set to be 1 year. Following this argument, Barbee et al. \cite{barbee2010comprehensive} considered a mission to a specific Apollo asteroid, and a possible transit trajectory with a duration of 160 days was identified. Assuming this number is applicable to the asteroid mining mission architecture, this allows for 15 days for the landing site identification and the landing itself, as well as 30 days of resource extraction operations. A conceptual sketch of the corresponding trajectory is shown in \prettyref{fig:proposed_trajectory}.

\subsection{Payload}
The size and mass of the spacecraft are limited by the mission top level requirements (T1). In this case, the biggest design driver of the spacecraft is the payload. We selected microwaves as the most appropriate water extraction technology for our purposes. The resulting main payload subsystems are therefore: microwave system, drilling system, anchoring system, prospecting system. In the following these subsystems are analyzed and defined in more detail.

\subsubsection{Microwave System}
Several experiments have been performed by Ethridge \cite{ethridge2016proposed} using Mars simulants and found extraction rates of 2 Wh/g, 7 Wh/g and 17 Wh/g of water, depending on the experimental conditions. One of the most well-known experiments positioned a microwave source inside a borehole made from simulant, in order to study the microwave reactions up to 1 m soil depth. Even at this depth, water extraction was still possible and successful. This method represents a lightweight and potentially low cost in-situ option.

\begin{figure}[htb]
 \includegraphics[width=0.6\textwidth]{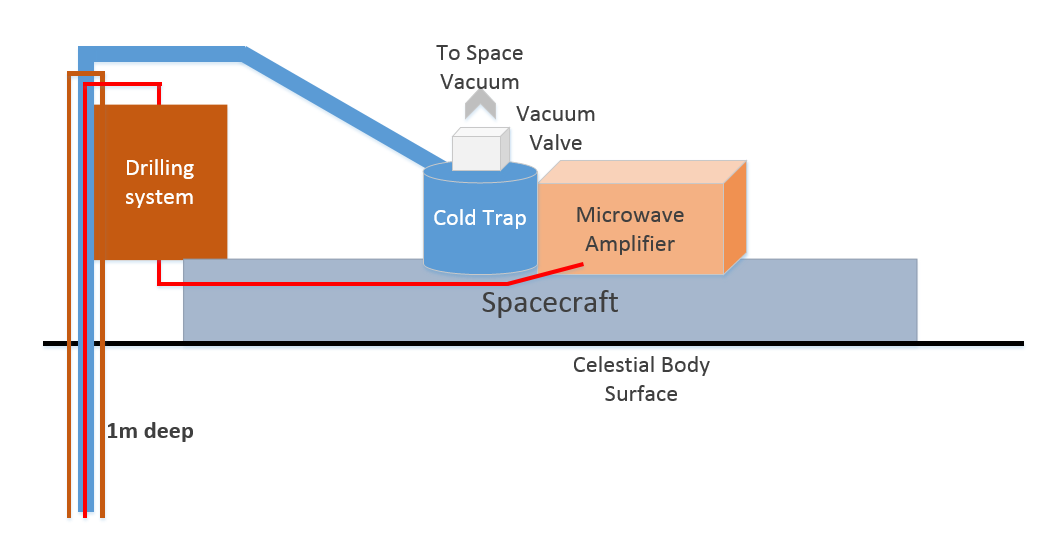}
 \caption{Schematic of a possible microwave extraction system, based on the system design in \cite{ethridge2016proposed}.}
 \label{fig:microwave_system2}
\end{figure}

A complete system for asteroid water extraction has also been proposed by Ethridge \cite{ethridge2016proposed} and a schematic is shown in \prettyref{fig:microwave_system2}.

The experiments were performed with a probe carrying a microwave source 1 meter deep from the top surface. Considering the same dimensions as a baseline case for the spacecraft payload, the components will be considered to fit a 1 meter design. Further proof of concept was provided at NASA MSFC by Ethridge and Kauler by demonstrating successfully the water extraction from cryogenic lunar permafrost simulant using microwaves \cite{ethridge2016proposed}. The experiment was repeated with a simulated carbonaceous chondrite asteroid. After exposure to microwaves, thermal decomposition occurs and released water vapour was captured in a cold trap. This recovered all water from the sample plus an additional 3 ml of water which had been chemically bound in the clay materials. For the proposed small spacecraft, the cold trap is also connected to the storage, fuel refinement and propellant tanks for refueling purposes.

\subsubsection{Drilling System}
Drilling in low strength materials such as plaster or limestone is feasible using commercially available drills, and these materials are likely to be representative of C-type asteroids \cite{badescu2013asteroids}. Therefore, it is assumed that similar conditions should be analyzed. 

1-meter class drills can fully fit into the payload envelope of a lander, which satisfies with T1 of the top level requirements. It is a compact drill that can be preassembled on Earth and does not need any robotic assembly. These drills can also be integrated with logging instruments to obtain in situ data while drilling. The data is critical if the drilling system is completely autonomous \cite{zacny2008drilling}. There are several drills that were manufactured and tested for extraterrestrial drilling such as the ones used by Zacny et al. \cite{zacny2013reaching} for Mars conditions at a Mars analog site in Antarctica, which offered a controlled environment in terms of temperature, pressure and formation properties. The equipment used consisted of the following components: drill head (consisting of two independent actuators with three different drilling modes - rotary, rotary-percussive and percussive), auger (1.2 m length, 25 mm diameter), z-stage ( moving the drill up to 1 m vertically and monitoring the force on the drill bit up to 100 N), electronics.

This system has the potential to be re-purposed for the asteroid mining mission, and allow a microwave probe to be inserted to a depth of up to 1 m. However, operation in a dusty vacuum will require very capable mechanical seals \cite{zacny2008drilling}.

\subsubsection{Anchoring System}
For an anchoring system to be both effective and low risk, it should react to all forces and torques caused by drilling operations.  It should also be able to release the spacecraft once operations have been completed.  

\begin{figure}[htb]
 \includegraphics[width=0.6\textwidth]{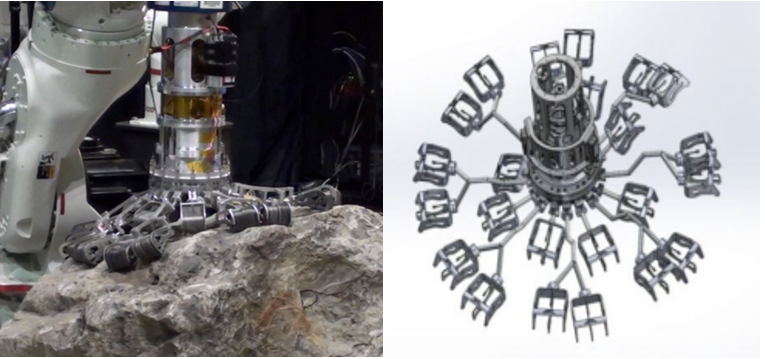}
 \caption{NASA's JPL microspine grippers \cite{parness2011}}
 \label{fig:JPL_grippers}
\end{figure}

Since it is expected that carbonaceous asteroids (Type C) have a rocky composition, a potential solution is the use of microspine grippers. These grippers use hundreds of tiny hooks to grip rough surfaces. This system provides support to counter forces in all directions away from the rock.  It can counter forces up to 160 N tangent to the rock surface, 150 N at 45$^\circ$, and 180 N normal to the surface.  These loads are greater than the expected drilling forces of 100 N \cite{Merriam2016}.  The gripping system concept is shown in \prettyref{fig:JPL_grippers}.

\subsubsection{Prospecting System}
While proper prospection is expected to have taken place before the mining missions, the spacecraft is envisioned to include instruments to validate and identify the target water resources in a short range as the spacecraft approaches the celestial body. Based on the prospection studies and proposed projects, optical sensors are used for prospections operations. More specifically, the optical payload identifies spin rate and size, radar determines the 3-dimensional shape and IR telescopes determine albedo and refines the type of asteroid \cite{hellgren2016asteroid}. Furthermore, using a thermal-IR imager and a near-IR spectrometer combined, it is possible to observe and estimate the geology and thermo-physical properties of an asteroid, allowing for the detection of organic and hydrated materials \cite{hellgren2016asteroid}. 

Bonin et al. \cite{bonin2016prospector} proposed a viable mid-infrared camera with 0.5 m spatial resolution at 10 km range and spectrometer in order to identify the asteroid characteristics for the DSI Prospector-1 spacecraft. Given this system's ability to characterize asteroid surface features and identify potential landing locations, it will be included in this project. However, cameras and star trackers are also used for attitude and control and therefore, they are considered part of that subsystem and not the payload.  

\subsection{Spacecraft Bus Definition}

To properly estimate the capabilities and economic cost/return of a miniaturized asteroid mining spacecraft, the propulsion, power, guidance, and communication systems are defined in more detail in the following sections.

\subsubsection{Propulsion}

\begin{wrapfigure}[18]{r}{0.4\textwidth}
	\includegraphics[width=\textwidth]{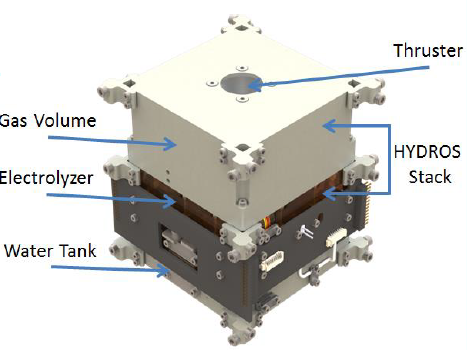}
	\caption{Architecture of the water electrolysis engine. 1U HYDROS\texttrademark thruster \cite{Karsten2016}.} 
	\label{fig:water_engine}
\end{wrapfigure}

Given the spacecraft requirements and the possibility to refuel at the target asteroid (top level requirement T3), water thrusters are the most viable option to satisfy those requirements. A schematic of a water electrolysis thruster system developed by Tethers Unlimited and NASA \cite{Karsten2016} is presented in \prettyref{fig:water_engine}. This system has an ISP $\leq$ 300 s.

The return transit phase starts once the spacecraft is refueled at the target asteroid. From there it needs to return the collected water to a useful orbit, and then start to the next asteroid. As such, a round trip was defined from the moment the spacecraft departs the asteroid until it reaches the next asteroid. Consequently, the amount of water that can be extracted with the on-board equipment is the real limit for the maximum travel distance.

Whether an asteroid is ``accessible” by the proposed propulsion system depends on a variety of factors including the orbit’s geometry, phasing, inclination, and amount of fuel available. The project assumes a suitable asteroid is selected in advance of each mining mission.

To estimate the required propellant mass, the required change in the spacecraft's velocity to reach the asteroid needs to be known. This velocity, also called $\Delta V$, measures the amount of energy required to go from one orbit to a another one. The most simple case of such an orbit change is called a ``Hohmann transfer", yielding a simple formula for the required $\Delta V$, which is described in more detail in \hyperref[app:AppendixB]{App. B}. The amount of fuel is then given, again in the most simple case, by the Tsiolkovsky rocket equation. For the computations in this manuscript, we will use these most simple methods. Of course, this technique can also be used in reverse, i.e. for a given amount of fuel, the possible orbit change can be determined.
In order to satisfy requirement T10, 5 kg are allocated to a small electric propulsion engine as a back-up option to recover the spacecraft in case of an emergency or failure of the main propulsion system. 

\subsubsection{Guidance, Navigation and Control }
From the mission phases in \prettyref{fig:proposed_architecture2}, the following operational modes are identified: orbit insertion, acquisition, nominal operation, Prospecting, slew, contingency, and landing/docking. 


A standard AOCS (Attitude and Orbit Control Systems) suite that meets the top level requirements can be used for this mission, and is comprised of 1x IMU (Inertial Measurement Unit), 3x Star trackers, 3x Reaction wheels, 3x Sun sensors, 1x Propulsion reaction control system, 2x Wide angle cameras, and 2x Altimeter. Such an AOCS suite was proposed with slight variations for the spacecraft prospectors of DSI (Deep Space Industries) \cite{bonin2016prospector} and the ESA MarcoPolo candidate mission \cite{chalex2013}. 


\subsubsection{Communication}
The communication challenges involve ground support for the mission, trajectory corrections, control in case of emergencies, and the transmission of data for analysis.  

For the spacecraft side, the Prospector-1 mission from DSI (Deep Space Industries) proposes the use of X-band communication. In their studies, they also define some link budget characteristics such as the transponders power, to ensure continuous communication \cite{bonin2016prospector}. Considering data volume, the prospection phase is the system driver, however, no real time communication is needed. For a similar application, the ESA MarcoPolo mission included three antennas for communications: a high gain antenna, a medium gain antenna, and two low gain antennas, basing their design on the Bepi Colombo and Solar Orbiter missions \cite{chalex2013}.
The current small spacecraft shall include a similar configuration in the X band, as the ESA Marco Polo mission and DSI spacecraft. 

\section{Results}

The estimation of mass and power budgets is an iterative process, as mass, power and propulsive capabilities are highly interdependent quantities. In the following, only the final results and their justifications are presented. Margins of 20\% have been added to all calculated values.

\subsection{Spacecraft Mass Budget and Range}
The weight of the spacecraft is highly dependent on the payload. Given these characteristics of the corresponding components, the payload mass, volume, and power estimates are shown in \prettyref{tab:payload-bg}, \hyperref[app:AppendixA]{App. A}.

The structure and the thermal subsystem masses are estimated from the remaining spacecraft dry-weight. Following standard techniques \cite{wertz2011space}, the complete mass budget is shown in \prettyref{tab:mass-bg}.

\begin{table}[htb]
\centering
\begin{tabular}{|l|c|l|}
\hline
\multicolumn{1}{|c|}{\textbf{Subsystem}} & \multicolumn{1}{c|}{\textbf{Weight [kg]}} & \multicolumn{1}{c|}{\textbf{Comments}} \\ \hline
Payload: Anchoring & 12 & \begin{tabular}[c]{@{}l@{}}A system of three grippers for anchoring\\with 120 degrees difference is preferred\end{tabular} \\ \hline
Payload: Drilling & 30 & \begin{tabular}[c]{@{}l@{}}Two drills are selected, a main drill and a \\redundant unit\end{tabular} \\ \hline
Payload: Extraction & 124 & \begin{tabular}[c]{@{}l@{}}8 microwave generators are preferred. 7  \\for operations + 1 as a redundant unit + \\water conduits and the water tank\end{tabular} \\ \hline
Payload: Prospection & 2 & One set of prospection sensors \\ \hline
Attitude control & 35.6 & Mass budget estimated \\ \hline
Communications & 9.7 & Mass budget estimated \\ \hline
\begin{tabular}[c]{@{}l@{}}Command and data \\handling \end{tabular} & 2 & Processor and electronics \\ \hline
Thermal & 9 & 5\% of the total mass \\ \hline
Power & 33.5 & Mass budget estimated \\ \hline
Structure & 30 & 15\% of the total mass \\ \hline
Propulsion & 25& Scaling the proved cubesat designs \\ \hline
Margin & 60& 20\% Margin due to new design \\ \hline
\textbf{Dry mass} & \textbf{372.8} & \textbf{Estimated Dry mass} \\ \hline
Propellant mass & 75 & \begin{tabular}[c]{@{}l@{}}Total amount of water required for 0.03  \\AU range + 10 \% margin\end{tabular} \\ \hline
\textbf{Wet mass} & \textbf{448} & \textbf{Estimated wet mass of the spacecraft} \\ \hline
\end{tabular}
\caption{Estimated spacecraft mass budget. The properties of the power system have been calculated iteratively, see details in the next section. }
\label{tab:mass-bg}
\end{table}

The requirement for a "small" spacecraft is set to 500 kg which limits mass we can consider when the spacecraft is launched from earth. Refueling the spacecraft at the asteroid during mining operations allows us to stay within this mass limit. Moreover, since the spacecraft intends to extract water as a payload, its "dry" mass will increase after the extraction of water at the target asteroid. Therefore, the wet mass calculations shown in \prettyref{tab:mass-bg} include only the water required for a ``one-way" trip to the asteroid. 

For the mission the spacecraft needs to complete a round trip with a water payload of 100 kg to a useful orbit and then back to the asteroid. Thus, the total amount of water required as propellant is estimated to be $\sim 150 kg$ for a maximum distance of $\approx 0.03$ AU and $\Delta V \approx 437$ m/s one-way. According to \prettyref{fig:water_pool}, this would give the spacecraft access to more than one million liters of water.

As a result, approximately 150 kg additional water has to be mined for refueling in addition to the 100 kg for commercial purposes, i.e. a total of $\sim 250$ kg of water has to be extracted. Considering the previously described microwave system, \prettyref{sec:MissionCraftReq}), with a water extraction efficiency of 2 Wh/g and standard 200W solid state amplifier power supplies, 7 microwave probes and power supplies are required to extract this amount of water within the given extraction duration of 30 days per mission with 15 net operational days, assuming occultation of the sun for 50\% of the time. 

\subsection{Spacecraft Power Budget}

The power budget is determined based on the amount of water that needs to be extracted and the technique used. The mining extraction period is defined as 30 days, see \prettyref{sec:MissionCraftReq}. It is assumed that, on average, the spacecraft operates in sunlight for half of that time period.  

The amount of power required by the water extraction system is 1400 W (7 microwave generators). Estimates of the total spacecraft power requirements can be performed based on the payload power requirements for small spacecraft from statistical data \cite{wertz2011space}. The resulting complete critical power budget is presented in \prettyref{tab:power-bg}. A more detailed investigation of the spacecraft power cycles during different mission cycles gives the total power system weight of 33.5 kg presented in \prettyref{tab:mass-bg}, including 9.5 $\mathrm{m}^2$ of solar arrays (Ultra flex), batteries (Lithium-ion), and other power control equipment \cite{wertz2011space}.

\begin{table}[h!]
	\centering
	\begin{tabular}{|l|c|c|c|}
		\hline
		\begin{tabular}[c]{@{}l@{}}\textbf{Spacecraft} \\  \textbf{Subsystem} \end{tabular} & \begin{tabular}[c]{@{}c@{}}\textbf{Percentage of}\\\textbf{operating}\\\textbf{power selected}\\\textbf{\%} \end{tabular} & \textbf{Top level requirement}&\begin{tabular}[c]{@{}c@{}}\textbf{Power} \\  \textbf{estimate}\\\textbf{W}\end{tabular} \\ \hline
		\textbf{Payload} & 50 & Power for water extraction (T2) & 1400 \\ \hline
		\textbf{Propulsion} & 5 & \begin{tabular}[c]{@{}l@{}}Estimation of the idle behavior  \\ including the backup propulsion \\ recovery system (electric) (T10)\end{tabular} & 60 \\ \hline
		\textbf{Attitude Control} & 10 & \begin{tabular}[c]{@{}l@{}}Including important sensors for  \\ cruising, prospecting and  \\ landing (T4)\end{tabular} & 120 \\ \hline
		\textbf{Communications} & 10 & \begin{tabular}[c]{@{}l@{}}Including sending data from 0.1   \\AU (T7)\end{tabular} & 120 \\ \hline
		\textbf{Command and DH} & 10 & \begin{tabular}[c]{@{}l@{}}Including data processing and \\high degree of autonomy (T6)\end{tabular} & 120 \\ \hline
		\textbf{Thermal control} & 5 &  \begin{tabular}[c]{@{}l@{}}Survivability through all mission \\phases (T11)\end{tabular} & 60 \\ \hline
		\textbf{Power} & 10 & \begin{tabular}[c]{@{}l@{}}Including battery charging \end{tabular} & 120 \\ \hline
		\textbf{Structure} & 0 & No power considered & 0 \\ \hline
		\textbf{Total} & 100 &- & 2000 \\ \hline
	\end{tabular}
	\caption{Proposed critical spacecraft power budget}
	\label{tab:power-bg}
\end{table}

\FloatBarrier
The most intensive power consumption occurs during extraction, in which the spacecraft is anchored to the asteroid. In such scenario, the attitude control and propulsion are not considered active and the thermal control power requirements would require half the power estimated. Since a 20\% margin is often considered for novel designs \cite{wertz2011space}, this is added to the extraction period, reaching a maximum power requirement of 2150 W. 

\subsection{Cost and Feasibility Analysis}

Parametric cost estimation is selected for the cost analysis as we perform the analysis on a concept description. The parametric method uses empirical correlations to go from a product description to its cost. The method is widely used when the cost of only a few key variables has to be estimated \cite{wertz2011space}. The most common example of parametric cost estimation is the Cost Estimating Relationship (CER), which is based on statistical models. CER derived from representative and validated data from past projects are expected to provide valid estimations of new mission concepts \cite{wertz2011space}. Since we are dealing with a spacecraft weighing less than 500 kg, the CER we are using are taken from the Small Spacecraft Cost Model (SSCM). The SSCM provides an average of the estimated cost, as it uses a database of small spacecraft missions. The first payload unit includes the Research, Development, Test and Evaluation (RDT\&E) costs.

\begin{table}[htb]
	\centering
	\begin{tabular}{|p{4cm}|p{6cm}|c|}
		\hline
		\multicolumn{1}{|c|}{\textbf{Cost Component}} & \multicolumn{1}{c|}{\textbf{CER Estimation Criteria}} & \multicolumn{1}{c|}{\textbf{\begin{tabular}[c]{@{}c@{}}RDT\&E (NRE) \\ Cost (FY10 \$K)\end{tabular}}} \\ \hline
		Extraction & Mass: 12 kg, Power: 200 W, TRL: 5 & 22600 \\ \hline
		Drilling & Mass: 15 kg,  Power: 200 W, TRL: 5 & 9100 \\ \hline
		Anchoring & Mass: 12 kg, Power: 20 W, TRL: 5 & 10700 \\ \hline
		Prospection & \begin{tabular}[c]{@{}l@{}}Mass: 2 kg, Power: 12 W,\\   Design life: 10 years\end{tabular} & 6400 \\ \hline
		\textbf{Payload total cost} &  & 48800 \\ \hline
	\end{tabular}
	\caption{SSCM cost estimation of the first payload unit}
	\label{tab:Table_CER_Payload}
\end{table}

Many of the cost estimations in the literature only refer to payloads which have already been launched mainly in communications and remote sensing. The payload proposed in this document does not belong to any of those categories. Therefore, the approximate cost of the payload is estimated based on heritage and similarity. 

The extraction system is estimated following the thermal subsystem CERs of the spacecraft bus. The reason for this selection is the payload similarity, i.e. the extraction system has the function to heat up part of the asteroid regolith and collect the water. The drilling system is estimated based on the structure and mechanisms subsystem CERs, as it is composed of mainly mechanical parts with the only objective to drill the asteroid surface. The anchoring system is also be estimated based on the structure and mechanisms subsystem CERs of the spacecraft bus given its similarity. The prospecting payload is estimated based on the optical planetary instruments CERs which includes cameras, spectrometers, interferometers and IR sensors. The results are given in \prettyref{tab:Table_CER_Payload} with a total cost of \$48.8 million.

\begin{table}[htb]
\centering
\begin{tabular}{|p{5cm}|p{5cm}|c|}
\hline
\multicolumn{1}{|c|}{\textbf{Cost Component}} & \multicolumn{1}{c|}{\textbf{CER Estimation Criteria}} & \multicolumn{1}{c|}{\textbf{\begin{tabular}[c]{@{}c@{}}RDT\&E (NRE) \\ Cost (FY10 \$K)\end{tabular}}} \\ \hline
\begin{tabular}[c]{@{}l@{}}Integration, Assembly and\\   Test (bus + payload)\end{tabular} & \begin{tabular}[c]{@{}l@{}}S/C Bus and Payload\\   Cost: \$113.6M\end{tabular} & 9000 \\ \hline
Launch Vehicle & SpaceX Falcon 9 Heavy \cite{FalconH} & 90000 \\ \hline
Program Level & S/C Bus Cost: \$64.8M & 14800 \\ \hline
Launch and Orbital Support & S/C Bus Cost: \$64.8M & 4000 \\ \hline
Ground Support Equipment & S/C Bus Cost: \$64.8M & 4300 \\ \hline
\end{tabular}
\caption{Estimation of additional costs. S/C = Spacecraft.}
\label{tab:Table_CER_Additional}
\end{table}

The estimated cost of the first spacecraft bus excluding the payload is \$64.8 million. A more precise breakdown of the estimate is given in \prettyref{tab:Table_CER_Bus}, \hyperref[app:AppendixA]{App. A}. 

Additional cost such as program level expenses, launch vehicle, integration and testing, are estimated using established guidelines \cite{wertz2011space}, see \prettyref{tab:Table_CER_Additional}. Finally, this approach provides an estimated total mission cost for the first unit, including flight software cost and ground support equipment which only has to be paid once, of \$145.7 million plus \$90 million for one Falcon Heavy launch vehicle. Since SpaceX gives a payload capability of $\sim 64$ metric tons into LEO \cite{FalconH}, launching only one of our mining spacecraft at a time would be a waste. Thus, in the subsequent analysis we will consider the optimal case of either fully using each Falcon Heavy's payload capability, launching multiple mining spacecraft at once, or using rideshare options.

\subsection{Business Case Analysis}
The cost estimation is used to perform an economic return analysis. In this part, we consider the mass production of the spacecraft described earlier. When several of the same units are produced the economy of scale principle applies, i.e. the cost per unit is reduced. The learning curve accounting for this effect is \cite{wertz2011space},
\begin{equation}
\mathrm{Total Lot Cost} = T_1 \, N^{\left( 1+ln(S)/ln(2) \right)},
\label{eq:learning}
\end{equation}
where $T_1$ is the theoretical first unit production cost, excluding software and ground support equipment costs. $N$ is the quantity (lot size), $S $ learning curve slope in decimal form. Here, a slope of 0.85 is used for cost calculations, which is a conservative estimate.

\begin{equation}
\mathrm{First Year Cost} = \mathrm{Total Lot Cost} + N_r C_r + C_s + C_g,
\end{equation}

$N_r$ are the number of SpaceX Falcon Heavy launch vehicles required, at a cost per launch vehicle of $C_r$ of  \$90 million \cite{FalconH}. We assume here that one single rocket can lift 144 spacecraft according to their previously estimated mass. Nonetheless, for a small number of smaller spacecraft a ridehsare option might be more affordable than booking out a full rocket. We consider rideshare options with a price point of \$3000/kg (similar to the price for bringing water into LEO, see text below and \prettyref{tab:economic-return}). Thus, rideshare is no longer advantageous once the cost per spacecraft falls below \$1.344 million, at which point we assume a full Falcon Heavy launcher is used. 

To estimate the price of water delivered to different orbits from earth, we consider different estimates given in \prettyref{tab:economic-return}. The current improvement and proliferation of launch vehicles could allow for a lower price per [kg] in the future. Flight software cost $C_s$ is \$27.5 million, and ground support equipment is \$4.3 million, see \prettyref{tab:Table_CER_Additional} and \prettyref{tab:Table_CER_Bus}. 

\begin{table}[htb]
\centering
\begin{tabular}{|p{6cm}|p{5.5cm}|}
	\hline
	\multicolumn{1}{|c|}{\textbf{Hardware and operation cost} } & \multicolumn{1}{c|}{\textbf{Cost/kg to certain orbit}} \\ \hline
	Production of first unit: \$113.6 M (from SSCM total cost) & Cost for 1 kg in LEO: \$3 K (average) \\ \hline
    Launcher: \$90 M (Falcon Heavy to LEO) & Cost for 1 kg in GTO: \$7.5 K (Falcon-9) \\ \hline
    Annual operations: \$5.7 M \;(average) & Cost for 1 kg in GSO: \$21.5 K (Proton-M) \\ \hline
     & Cost for 1 kg in Cis-lunar space: \$35 K  \\ \hline
\end{tabular}
\caption{Parameters and assumptions for the economic return analysis \cite{wertz2011space,FalconH,spacereport}.}
\label{tab:economic-return}
\end{table}

With the defined parameters, and applying the learning curve in \prettyref{eq:learning}, economic return graphs can be generated. However, the possibility for improvements or problems in launch capability, manufacturing and technology add uncertainty to any estimate we make. We perform a sensitivity and uncertainty analysis in which we vary certain input parameters to quantify this issue. Specifically, we assume normal distributions for the values of the learning curve slope ($S$), the cost of the first spacecraft ($T_1$), a scaling factor for the water price ($\alpha_{H_2O}$), and the launch capability per rocket ($m_{LPL}$, where $LPL$ = launcher payload) and we evaluate 3000 independent sample combinations to ensure clear trends and statistically converged results. The water price scaling factor is multiplied to the values given in \prettyref{tab:economic-return} to account for possible variations in the price point of water, i.e. a scaling factor of $\alpha_{H_2O} = 1.1$ indicates a cost increase of 10\%. It is important to note that a variation in the price of water is likely to be tied to variations in launch cost. Thus, the launch cost is not treated as an independent variable but multiplied by the same value of $\alpha_{H_2O}$ as the water price to account for this relationship. The utilized distributions are completely defined by their mean and standard deviation as summarised in \prettyref{tab:distribution-parameters}. The given standard deviations reflect what the authors perceive as realistic, possible variations in the corresponding parameters based on literature and current trends. 

\begin{table}[htb]
\centering
\begin{tabular}{|p{3cm}|p{3cm}|p{5cm}|}
	\hline
	\multicolumn{1}{|c|}{\textbf{Parameter}} & \multicolumn{1}{c|}{\textbf{Mean}} & \multicolumn{1}{|c|}{\textbf{Standard Deviation} }\\ \hline
    \multicolumn{1}{|c|}{$S$} & 0.85 & 0.025 \\ \hline
    \multicolumn{1}{|c|}{$T_1$} & \$113.6M & \$10M \\ \hline
    \multicolumn{1}{|c|}{$\alpha_{H_2O}$} & 1 & 0.1 \\ \hline
    \multicolumn{1}{|c|}{$m_{LPL}$} & 64,000 kg & 10,000 kg\\ \hline
\end{tabular}
\caption{Parameter space used for the normal distributions in the sensitivity analysis.}
\label{tab:distribution-parameters}
\end{table}

The sensitivity analysis reveals that the total cost of the proposed asteroid mining architecture is not dependent on $m_{LPL}$. The dependence on $T_1$ is linear but becomes weaker the more spacecraft are utilized, as would be expected from the learning curve approach which makes each mining spacecraft manufactured after the first one cheaper and cheaper. The learning curve slope $S$ is a strong driver of the overall cost and the dependence becomes more non-linear the more spacecraft are manufactured, suggesting that an ideal intermediate number of spacecraft exists at which a break-even point can be reached within a reasonable time frame while avoiding a non-linear cost increase, in case the learning effect is not as strong as hoped for. Finally, the scaling of the total cost with $\alpha_{H_2O}$ is rather interesting and will be investigated together with the revenue in the next paragraph.

\begin{figure}[htb]
	\centering
	\includegraphics[width=0.7\textwidth]{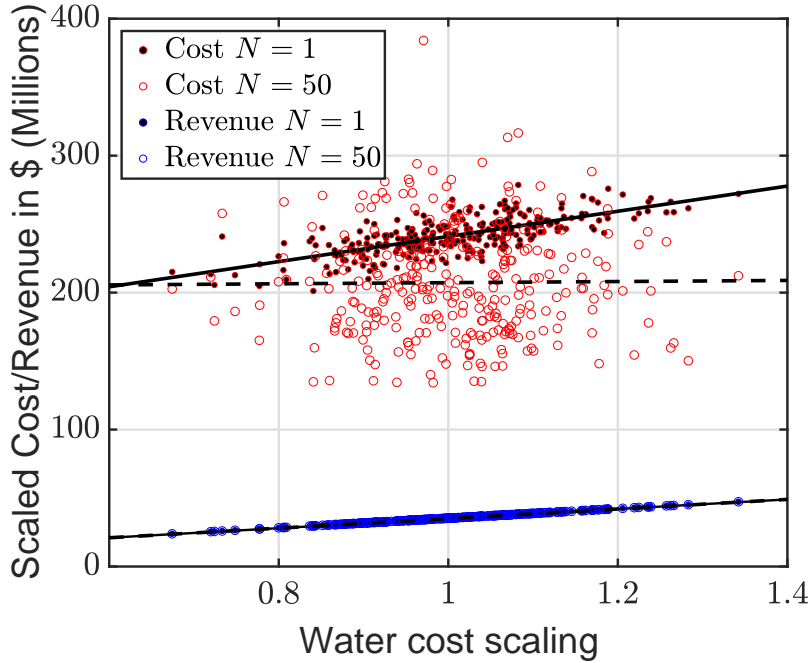}
	\caption{Shown is the sensitivity of cost and revenue to the water cost scaling factor. For clarity only every 10th data point out of 3000 sample combinations is shown. Cost and revenue values for $N=50$ are scaled to allow for proper comparison with the $N=1$ data. Black lines denote linear curve-fits for trend analysis: full lines correspond to $N=1$, dashed lines correspond to $N=50$.}
	\label{fig:sensitivity-learning-factor}
\end{figure}

As one would expect, the overall revenue of the mining venture is independent of all investigated parameters expect the water price scaling parameter $\alpha_{H_2O}$. The revenue and cost trends with $\alpha_{H_2O}$ are shown together in \prettyref{fig:sensitivity-learning-factor}. The water revenue (blue symbols) increases with $\alpha_{H_2O}$ in the same linear fashion no matter how many spacecraft are used. Based on our analysis, however, an increase in water prices also implies and increase in launch cost. For a single spacecraft the result is that the total cost (red symbols) of the venture increases as the water price increases, somewhat negating possible additional profits. However, using multiple mining spacecraft (e.g. $N=50$ but the trend holds for larger numbers) this relationship is weakening significantly, as indicated by the black trend lines. Therefore, using a larger number of spacecraft, given the proposed mission architecture and assumptions, the coupling of the total cost to launch price fluctuations is weak. This makes the approach more robust towards unexpected changes and allows one to maximize profits in case water prices are rising. Of course, this also implies that falling water prices would reduce profits while costs would change only slightly.     

\begin{figure}[h!]
	\centering
	\begin{subfigure}[t]{0.47\textwidth}
		\centering
		\includegraphics[width=\textwidth]{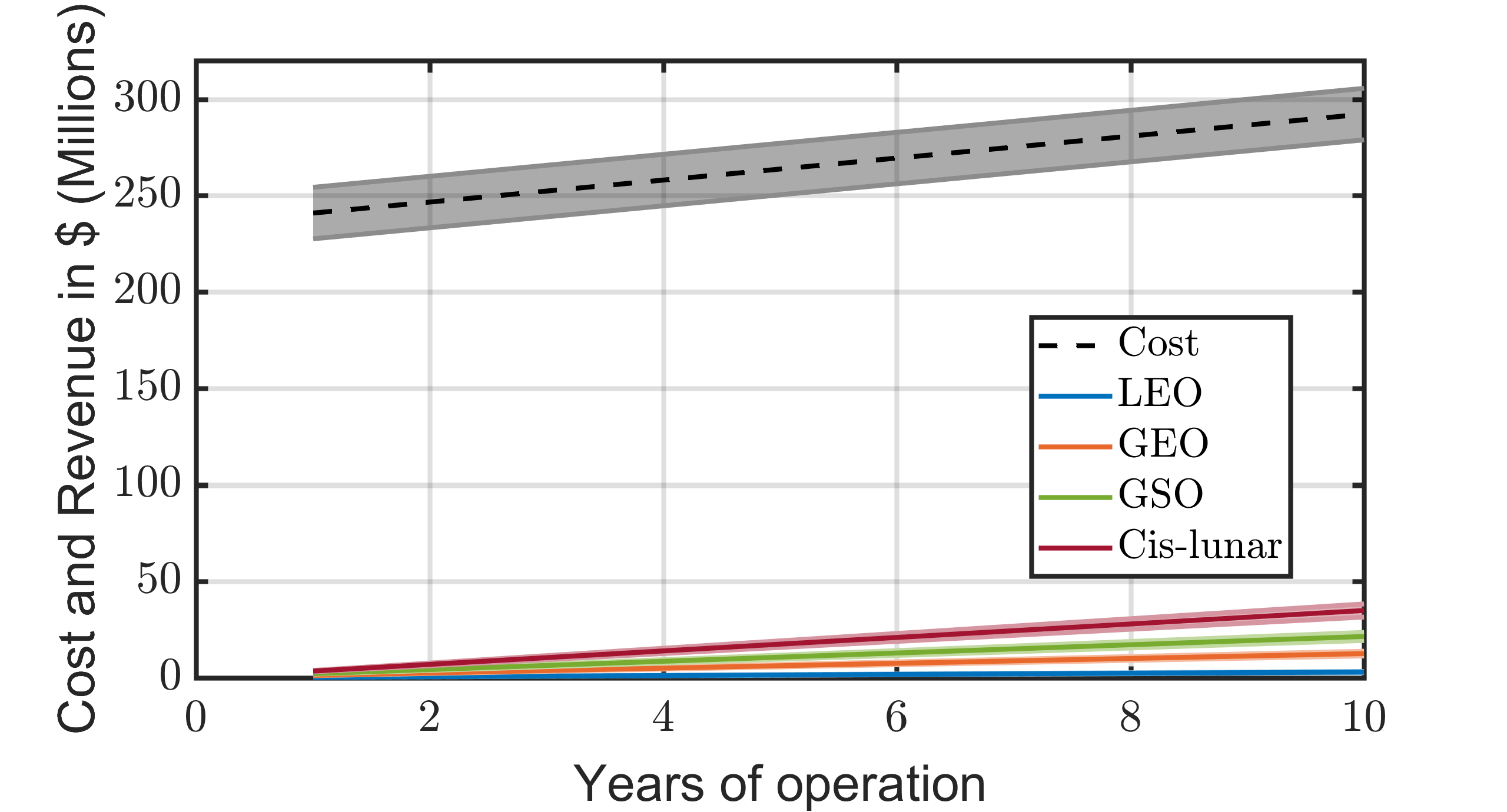}
		\caption{One water mining spacecraft.}
	\end{subfigure}%
	\hfill
	\begin{subfigure}[t]{0.47\textwidth}
		\centering
		\includegraphics[width=\textwidth]{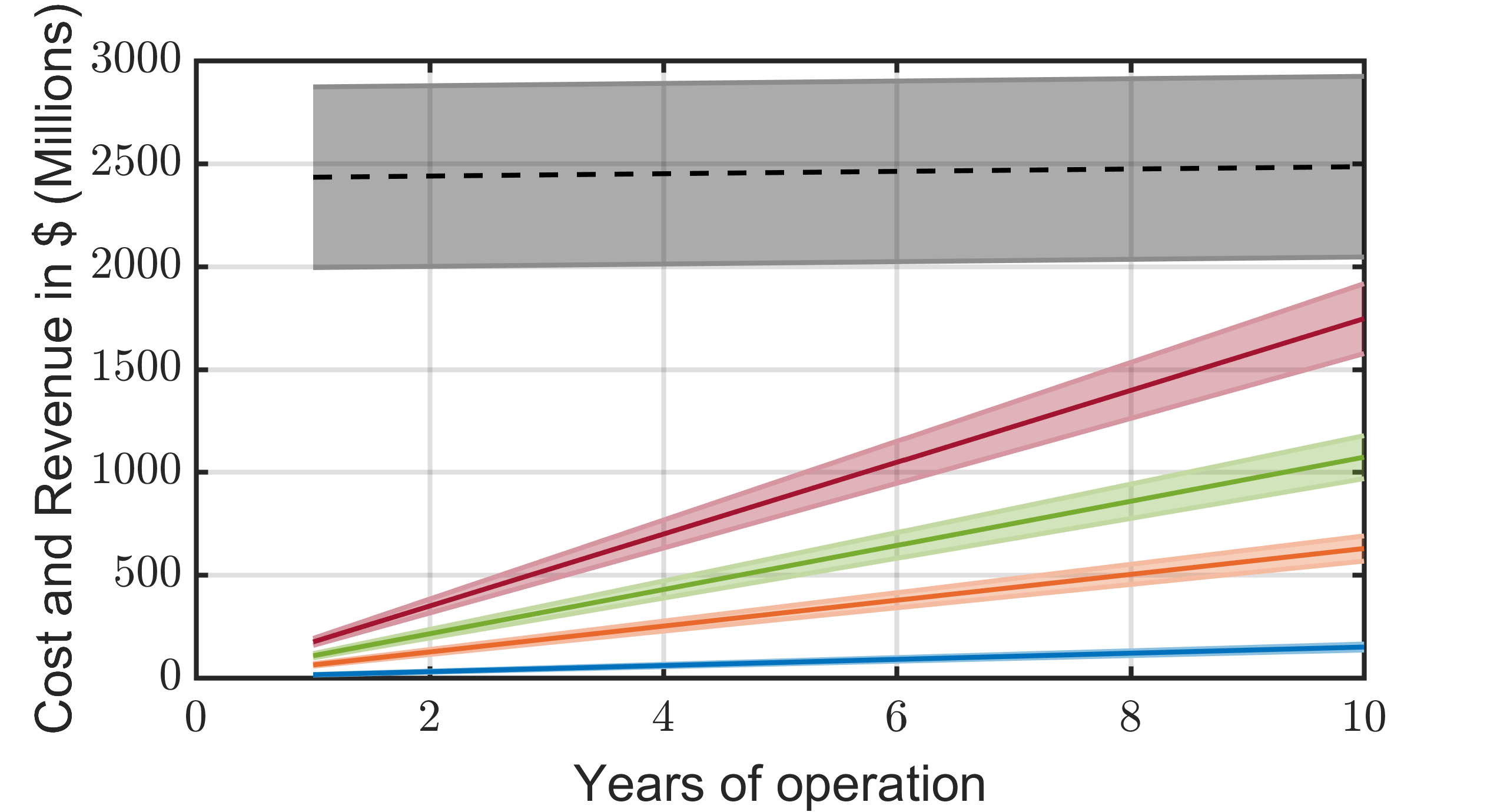}
		\caption{50 water mining spacecraft.}
	\end{subfigure}

	\begin{subfigure}[t]{0.47\textwidth}
		\centering
		\includegraphics[width=\textwidth]{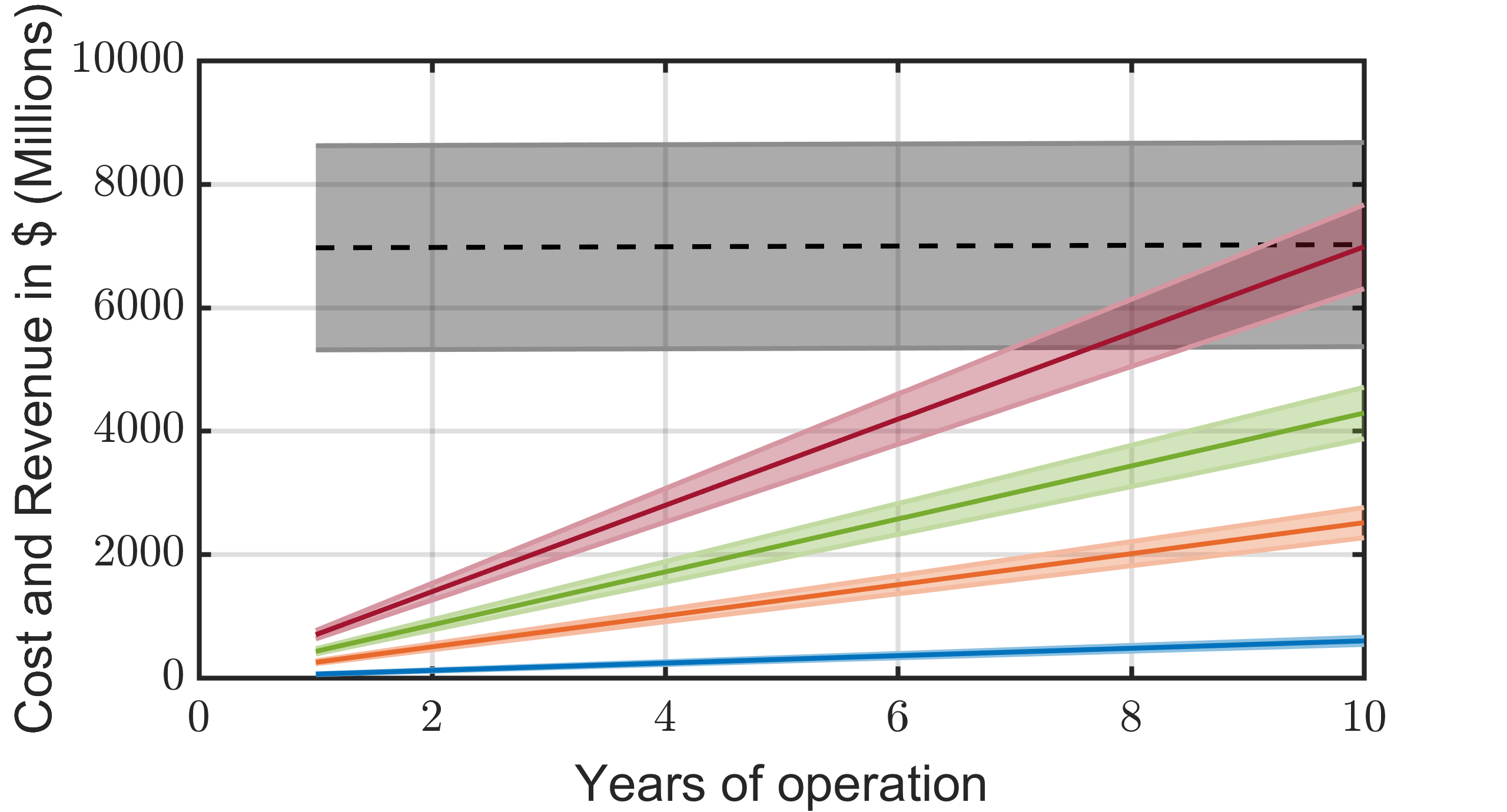}
		\caption{200 water mining spacecraft.}
	\end{subfigure}%
	\hfill 
	\begin{subfigure}[t]{0.47\textwidth}
		\centering
		\includegraphics[width=\textwidth]{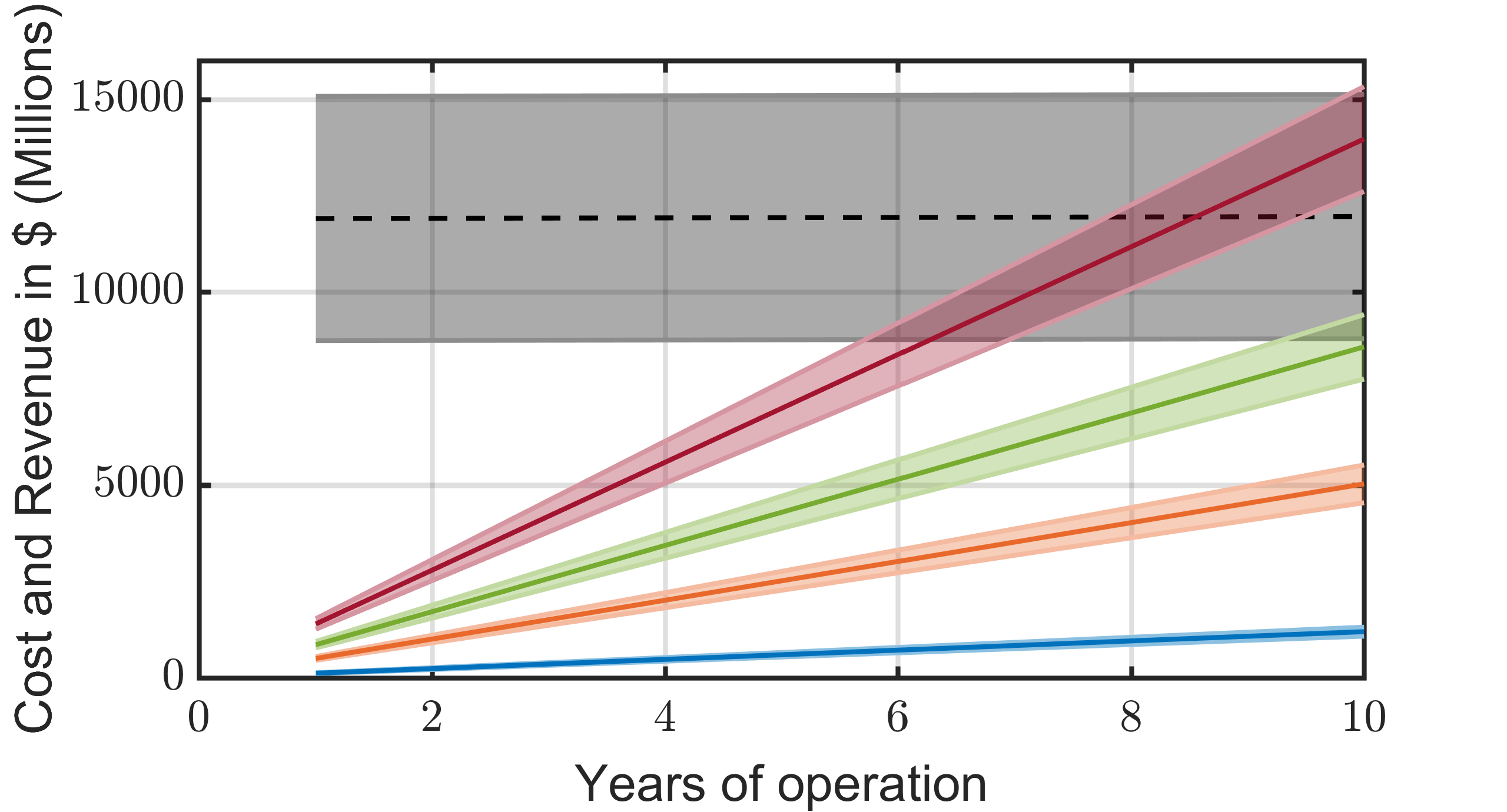}
		\caption{400 water mining spacecraft.}
	\end{subfigure}

	\caption{Cost analysis and economic return for (a) one, (b) 50, (c) 200, and (d) 400 spacecraft. The dashed black line indicates the spacecraft cost, including development, construction, launch, and annual operations cost.  Shadowed regions indicate 95\% confidence intervals based on the parameter space described in \prettyref{tab:distribution-parameters}. The revenue by sale and usage of the mined water is shown for LEO (blue line), GTO (orange line), GSO (green line), and Cis-lunar space (red line).}
 	\label{fig:cost_analysis_summary}
\end{figure}

Now, we evaluate the overall economic feasibility of the presented asteroid mining architecture and cost model. The resulting cost-revenue graphs with 95\% confidence intervals for 3000 independent samples of the above described parameters are presented in \prettyref{fig:cost_analysis_summary} utilizing one, 50, 200, and 400 mining spacecraft.

The first break-even point occurs for 200 produced spacecraft and utilization of the returned water in cis-lunar space, as shown in \prettyref{fig:cost_analysis_summary} (c). Profit is gained after approximately 10 years of operation, assuming no major failures. For even larger numbers of spacecraft ($\sim$400 in \prettyref{fig:cost_analysis_summary} (d)), geo-synchronous orbits also become interesting, while profit in cis-lunar space is more likely to occur after a shorter operational time, i.e. roughly 8 years. Thus, the most likely scenario making asteroid mining economically feasible is utilization and sale of water in cis-lunar space, employing mass produced and highly reliable mining spacecraft. Mass production lowers the overall cost and scales up water availability. It is important to highlight, that several spacecraft can be launched by the use of only one heavy launch vehicle. Thus, making the overall required investment relatively insensitive to launch cost. Interestingly, even within the narrow band of variation considered (see \prettyref{tab:distribution-parameters}), the uncertainties of the results grow relatively large for larger number of mining spacecraft. Accordingly, for extremely favorable parameter combinations 200 spacecraft could achieve a break-even point after as little as 7 years of selling water in cis-lunar space (\prettyref{fig:cost_analysis_summary} (c)). For 400 spacecraft the earliest break-even point could be less than 6 years and even utilizing only 50 spacecraft can get close to being profitable, considering the uncertainty bands in \prettyref{fig:cost_analysis_summary} (d) and (b), respectively. This also highlights the need for further refinement of cost and return estimations for asteroid mining, to reduce the uncertainties in the outcome of the investment.

\FloatBarrier

\section{Conclusion and Further Discussion}

For the presented asteroid mining architecture, utilizing spacecraft designed to be below 500 kg in weight, the maximum distance to asteroid rendezvous from LEO is approximately 0.03 AU. From the corresponding delta-V of 437 m/s the NEAs that can be reached contain more than one million liters of water.

The economic analysis shows that using swarms of smaller spacecraft around 200 units are required to achieve an economically feasible operation within 10 years of operation. The concept has the advantages that it allows for rapid scaling up of mining operations and implements redundancy on the system level. Even for the 200 spacecraft fleet required to reach break-even in less than 10 years the up-front investment of $\approx \$7$ billion is below major acquisitions currently happening in our terrestrial economy, e.g. Amazon bought Whole Foods for \$13.7 billion in 2017 and the Vision Fund acquired \$93 billion during their 2017 funding round. Finally, the sensitivity analysis has shown that the mass-produced spacecraft cost is only weakly dependent on rising launch costs, thus allowing for maximization of profits from potentially rising water prices. On the other hand, the concept does not become profitable in less than 8 years and does not significantly profit from decreasing launch prices, within the parameter space explored. 

The sensitivity analysis also indicates that using a very large number of spacecraft can lead to a non-linear increase in overall cost for unexpectedly low benefits from mass production (as embodied by the learning curve slope parameter $S$). Therefore, an intermediate number of spacecraft seems a ideal, at which break-even is possible within 10 years but the cost dependence on $S$ is still linear. In the work presented, this number lies between 200 and 350 spacecraft.

Lowering the launch cost by using rideshare opportunities does not make the proposed mission architecture profitable within 10 years for small numbers of spacecraft. This assumes that ridesharing should not be pursued once its cost per spacecraft exceeds the cost of booking a full heavy launcher.

With regard to target markets, the analysis identified cis-lunar space as the use-case most likely to make steroid water mining economically feasible. Customers in this sector include lunar and lunar-orbit bases, transiting Astronauts, so-called deep space gateways (space stations), and operations in the L2 Lagrange point of the Earth-Moon system.

The analysis makes it obvious that further development and research is required to make asteroid mining more attractive for investors and more likely to succeed in general. The authors identified several areas that, if improved, would contribute to this goal: 

\begin{itemize}
  \item Further miniaturization of spacecraft components will reduce component and launch cost significantly for a large number of spacecraft.
 \item Water extraction techniques need to be explored further and qualified for usage with actual asteroid material. Water needs to be extracted in an efficient and effective manner, as that directly impacts spacecraft power requirements and the amount of water that can be extracted in a given time frame.
 \item Water and water-derived fuel propulsion systems need to reach higher efficiency levels and achieve a sufficient TRL.
 \item Much larger, monolithic mining spacecraft, including asteroid capturing concepts, represent alternative mission architectures for asteroid mining. Inter-comparison of economic feasibility between small, medium, and large spacecraft concepts will provide important information regarding the most promising path to take in the future.
 
\end{itemize}

\section*{Acknowledgments}
The authors would like to thank Dr. Lee Wilson and Dr. Andreas Hein for valuable discussions and their input to the presented work. Moreover, we would like to thank Dr. Andreas Hein for bringing up the idea that led to this study.

\newpage

\pdfbookmark{Appendix A - Additional Tabulated Information}{app:AppendixA}
\section*{Appendix A - Additional Tables}\label{app:AppendixA}

\begin{table}[htb]
	\centering
	\resizebox{\textwidth}{!}{%
		\begin{tabular}{|l|l|l|c|c|c|}
			\hline
			\textbf{\begin{tabular}[c]{@{}l@{}}Water extrac-\\ tion technique\end{tabular}} & \textbf{Scalability} & \textbf{Complexity} & \textbf{T. Feasibility} & \textbf{Durability} & \textbf{\begin{tabular}[c]{@{}c@{}}Total\\ Score\end{tabular}} \\ \hline
			\textbf{\begin{tabular}[c]{@{}l@{}}Vacuum drying\\ (Pneumatic Sys-\\ tem)\end{tabular}} & 3 & 3 & 7 & 7 & 27.7 \\ \hline
			\textbf{\begin{tabular}[c]{@{}l@{}}Hot air or steam\\ drying\end{tabular}} & 5 & 5 & 8 & 5 & 32 \\ \hline
			\textbf{\begin{tabular}[c]{@{}l@{}}Solar drying (fo-\\ cused light)\end{tabular}} & 3 & 6 & 10 & 9 & 38.4 \\ \hline
			\textbf{\begin{tabular}[c]{@{}l@{}}Inclined pipes\\ heating\end{tabular}} & 7 & 6 & 8 & 4 & 34.9 \\ \hline
			\textbf{\begin{tabular}[c]{@{}l@{}}Kettle / Pot\\ heating\end{tabular}} & 6 & 8 & 3 & 8 & 33.5 \\ \hline
			\textbf{Sifter heating} & 7 & 8 & 4 & 5 & 32.6 \\ \hline
			\textbf{Funnel heating} & 7 & 7 & 4 & 5 & 31.4 \\ \hline
			\textbf{\begin{tabular}[c]{@{}l@{}}Conveyor belt\\ (drum drying)\end{tabular}} & 4 & 5 & 3 & 5 & 23 \\ \hline
			\textbf{\begin{tabular}[c]{@{}l@{}}Microwave\\ drying\end{tabular}} & 8 & 6 & 9 & 7 & 41.8 \\ \hline
			\textbf{\begin{tabular}[c]{@{}l@{}}Capturing and \\ heating\end{tabular}} & 1 & 2 & 10 & 9 & 30.6 \\ \hline
		\end{tabular}%
	}
	\caption{Decision Matrix for the water extraction technique.}
	\label{tab:decision-mat}
\end{table}

\begin{table}[htb]
	\centering
	\begin{tabular}{|l|c|c|c|}
		\hline
		\multicolumn{1}{|c|}{\textbf{Component}} & \multicolumn{1}{c|}{\textbf{Mass}} & \multicolumn{1}{c|}{\textbf{Dimensions}} &\multicolumn{1}{|c|}{\textbf{Power}} \\ \hline
		\begin{tabular}[c]{@{}l@{}}\textbf{Magnetron or solid} \\  \textbf{state amplifier +} \\\textbf{waveguide}\end{tabular} & 12 kg & \begin{tabular}[c]{@{}l@{}}480 mm x 440  \\  mm x 130 mm\end{tabular} &200W \\ \hline
		\begin{tabular}[c]{@{}l@{}}\textbf{Water conduit +} \\  \textbf{cold trap }\end{tabular} & 1 kg &  \begin{tabular}[c]{@{}l@{}}140 mm D x   \\  160 mm H\end{tabular} &0\\ \hline
		\begin{tabular}[c]{@{}l@{}}\textbf{Drill + mechanical} \\   \textbf{components}\end{tabular} & 15 kg & \begin{tabular}[c]{@{}l@{}}1460 mm long x    \\  13 mm diameter H\end{tabular} &100-200W \\ \hline
		\textbf{Water storage tank} & 10 kg (empty)& \begin{tabular}[c]{@{}l@{}}970 mm H x   \\  420 mm D\end{tabular} &0 \\ \hline
		\begin{tabular}[c]{@{}l@{}}\textbf{Microspine grippers}  \\   \textbf{system}\end{tabular} & 4 kg & \begin{tabular}[c]{@{}l@{}}250 mm D x    \\  300 mm H \end{tabular} &20W \\ \hline
		\textbf{Prospection system} & 2 kg & - & 12W \\ \hline
	\end{tabular}
	\caption{Proposed spacecraft payload budget based on information available in the referenced literature.}
	\label{tab:payload-bg}
\end{table}

\begin{table}[htb]
	\centering
	\begin{tabular}{|l|l|r|}
		\hline
		\multicolumn{1}{|c|}{\textbf{Cost Component}} & \multicolumn{1}{c|}{\textbf{CER Estimation Criteria}} & \multicolumn{1}{c|}{\textbf{\begin{tabular}[c]{@{}c@{}}RDT\&E (NRE) \\ Cost (FY10 \$K)\end{tabular}}} \\ \hline
		Structure & Mass: 30 kg & 2400 \\ \hline
		Thermal & Mass: 9 kg & 800 \\ \hline
		\begin{tabular}[c]{@{}l@{}}Attitude determination and\\   control system\end{tabular} & Mass: 35.6 kg & 16700 \\ \hline
		Electrical Power system & Mass: 33.5 kg & 8000 \\ \hline
		Propulsion & \begin{tabular}[c]{@{}l@{}}S/C Dry Mass: 372.8 kg,\\   TRL: 5\end{tabular} & 6900 \\ \hline
		\begin{tabular}[c]{@{}l@{}}Telemetry, Tracking and\\   Command\end{tabular} & Mass: 9.7 kg & 1700 \\ \hline
		Command and Data Handling & Mass: 2 kg & 900 \\ \hline
		Flight Software & Lines of code: 50,000 & 27500 \\ \hline
		\textbf{Spacecraft Bus total cost} & \textbf{} & \textbf{64800} \\ \hline
	\end{tabular}
	\caption{Cost estimation for the spacecraft bus of the first unit.}
	\label{tab:Table_CER_Bus}
\end{table}

\FloatBarrier
\pdfbookmark{Appendix B - Hohmann transfer}{app:AppendixB}
\section*{Appendix B: Hohmann transfer}\label{app:AppendixB}
Whether an asteroid is “accessible” depends on a variety of factors including the orbit’s geometry, phasing and inclination. However, the project assumes a suitable asteroid to be selected in advance of each mining mission. Thus, only a series of estimations for $\Delta$V requirements regarding the asteroid distance will be considered. The least $\Delta$V needed for a transfer between two circular orbits is achieved by using a double-tangent transfer ellipse, also called a Hohmann transfer. The scenario considered is displayed in figure 9.
\begin{figure}[htb]
 \includegraphics[width=1\textwidth]{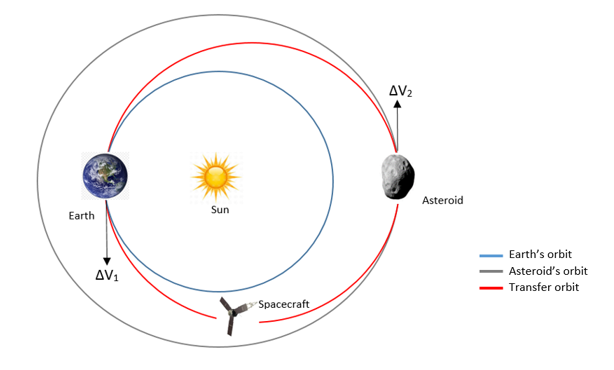}
 \caption{Hohmann transfer orbit for the asteroid rendezvous.}
 \label{fig:Hohmann}
\end{figure}

The method involves the firing at Earth's perihelion $\Delta \, V_1$, in order to accelerate the spacecraft, and for the asteroid capture, a second firing is imperative at the aphelion of the transfer orbit, $\Delta \, V_2$.
The Hohman equations for the $\Delta$V estimations from the heliocentric reference are:
\begin{equation}
\Delta V_1 = \sqrt{\frac{\mu_S}{r_T}}\,\Big(\sqrt{\frac{2r_A}{r_T+r_A}}-1\Big)
\end{equation}
\begin{equation}
\Delta V_2 = \sqrt{\frac{\mu_S}{r_A}}\,\Big(1-\sqrt{\frac{2r_T}{r_T+r_A}}\Big)
\end{equation}
Where: 
\\$\mu_S$ is the Sun’s gravitational parameter $= 1.327 \times 1020$ [$m^3$/$s^2$]\\
	$r_T$ is the Earth’s distance from the sun $= 1 AU = 1.496 \times 10^{11}$ [$m$]\\
	$r_A$ is the asteroid’s distance from the sun\\

The total change in velocity is equal to:
\begin{equation}
\Delta V = |\Delta V_1|+|\Delta V_2|
\end{equation}

\FloatBarrier
\section*{Author Vitae}

\subsection*{Pablo Calla}
\begin{wrapfigure}{r}{0.18\textwidth}
	\vspace{-0.5cm}
	\includegraphics[width=\textwidth]{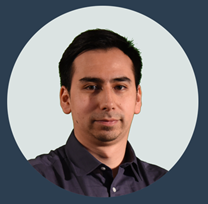}
\end{wrapfigure}

Pablo Calla is an electronic engineer who has specialized in renewable energy generation control systems engineering at the Universidad Mayor de San Andrés (UMSA) in Bolivia. He also holds a Master of Science in space studies from the International University (ISU) in France. He was selected among other engineers from Bolivia to be part of the development of the Bolivian Government’s TKSat-1 satellite project in China. Pablo is a cofounder and CTO of Maana Electric SA, where he is responsible for technology development.

\newpage

\subsection*{Dan Fries}

\begin{wrapfigure}{r}{0.18\textwidth}
	\vspace{-0.5cm}
	\includegraphics[width=\textwidth]{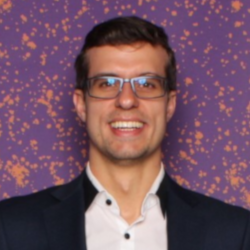}
\end{wrapfigure}

Dan Fries is a PhD candidate in Aerospace Engineering at the Georgia Institute of Technology, USA, working on supersonic combustion and mixing. He also holds Master's degrees in Aerospace Engineering from both Georgia Tech and the University of Stuttgart, Germany. Dan is a member of the Technical Committee of the Initiative for Interstellar Studies and has a strong background in propulsion physics and space systems engineering. He has participated in multiple, international projects and competitions aiming at the development of novel space mission architectures.

\subsection*{Prof. Chris Welch}

\begin{wrapfigure}{r}{0.18\textwidth}
	\vspace{-0.5cm}
	\includegraphics[width=\textwidth]{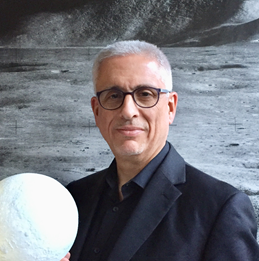}
\end{wrapfigure}

Chris Welch is Professor of Space Engineering and Director of Master’s Programs at the International Space University in Strasbourg, France. He has a PhD in Spacecraft Engineering (Cranfield University), an MSc in Experimental Space Physics (University of Leicester) and a BSc in Physics (Cardiff University). Chris is Vice-President of the International Astronautical Federation and a former member of European Commission H2020 Space Advisory Committee. He is also a Fellow of the British Interplanetary Society (BIS), the Royal Aeronautical Society and the Royal Astronomical Society, as well as a member of the International Academy of Astronautics Small Satellite Committee.

\FloatBarrier

\bibliographystyle{elsarticle-num}
\bibliography{LowCostAsteroidMining_Lit}

\begin{thebibliography}{10}
\expandafter\ifx\csname url\endcsname\relax
  \def\url#1{\texttt{#1}}\fi
\expandafter\ifx\csname urlprefix\endcsname\relax\def\urlprefix{URL }\fi
\expandafter\ifx\csname href\endcsname\relax
  \def\href#1#2{#2} \def\path#1{#1}\fi

\bibitem{badescu2013asteroids}
V.~Badescu, Asteroids: Prospective energy and material resources, Springer
  Science \& Business Media, 2013.

\bibitem{hellgren2016asteroid}
V.~Hellgren, Asteroid mining: a review of methods and aspects, B.{S}. thesis,
  Lund University, NGEK01 20161 (2016).

\bibitem{sanchez2011asteroid}
J.~Sanchez, C.~McInnes, Asteroid resource map for near-earth space, Journal of
  Spacecraft and Rockets 48~(1) (2011) 153--165.

\bibitem{ASTRA2010}
G.~Alotaibi, J.~Boileau, H.~Bradshaw, B.~Criger, R.~Chalex, J.~Chun,
  D.~Desjardins, J.~Dhar, A.~Artiles, K.~Ellis, {ASTRA}: Asteroid mining
  technologies roadmap and applications, International Space University (2010)
  1--100.

\bibitem{sanchez2012assessment}
J.~P. Sanchez, C.~R. McInnes, Assessment on the feasibility of future
  shepherding of asteroid resources, Acta Astronautica 73 (2012) 49--66.

\bibitem{JPL-CNEOS}
Center for near-earth object studies, \url{https://cneos.jpl.nasa.gov/},
  accessed: 2017-06-15.

\bibitem{PlanetRes}
Arkyd-301 mission by {P}lanetary {R}esources,
  \url{https://www.planetaryresources.com/missions/arkyd-301/}, accessed:
  2017-06-10.

\bibitem{zacny2013asteroid}
K.~Zacny, P.~Chu, J.~Craft, M.~M. Cohen, W.~W. James, B.~Hilscher, Asteroid
  mining, in: Proceedings of the AIAA SPACE 2013 conference and exposition,
  2013, pp. 10--12.

\bibitem{studies2012asteroid}
J.~Brophy, F.~Culick, L.~Friedman, P.~Llanos, et~al., Asteroid retrieval
  feasibility study, Tech. rep., Keck Institute for Space Studies, California
  Institute of Technology (2012).

\bibitem{dula2015space}
A.~M. Dula, Z.~Zhenjun, Space mineral resources: A global assessment of the
  challenges and opportunities, Tech. rep., Study conducted under the auspices
  of the {I}nternational {A}cademy of {A}stronautics ({IAA}) (2015).

\bibitem{ross2001}
S.~Ross, Near-earth asteroid mining, Tech. rep., {I}nternational {S}pace
  {U}niversity (2001).

\bibitem{rivkin2019many}
A.~S. Rivkin, F.~E. DeMeo, How many hydrated {NEO}s are there?, Journal of
  Geophysical Research: Planets 124~(1) (2019) 128--142.

\bibitem{wiens2001water}
J.~Wiens, F.~Bommarito, E.~Blumenstein, M.~Ellsworth, T.~Cisar, B.~McKinney,
  B.~Knecht, Water extraction from martian soil, in: Fourth Annual {HEDS-UP}
  Forum, 2001.

\bibitem{bernold2013}
L.~Bernold, Asteroids, Springer, 2013, Ch. Closed-Cycle Pneumatics for Asteroid
  Regolith Mining, pp. 345--364.

\bibitem{sercel2016}
J.~Sercel, Asteroid provided in-situ supplies ({APIS}): A breakthrough to
  enable an affordable {NASA} program of human exploration and commercial space
  industrialization, Tech. rep., {NASA} Innovative Advanced Concepts ({NIAC})
  (2016).

\bibitem{larson1992space}
W.~J. Larson, J.~R. Wertz, Space mission analysis and design, Microcosm Press,
  2005.

\bibitem{agasid2015}
E.~Agasid, R.~Burton, R.~Carlino, G.~Defouw, A.~D. Perez, A.~G.
  Karacalio\u{g}lu, B.~Klamm, A.~Rademacher, J.~Schalkwyck, R.~Shimmin,
  J.~Tilles, S.~Weston, Small spacecraft technology state of the art, Tech.
  rep., NASA {A}mes {M}ission {D}esign {D}ivision (2015).

\bibitem{zacny2013reaching}
K.~Zacny, G.~Paulsen, C.~McKay, B.~Glass, A.~Dav{\'e}, A.~Davila, M.~Marinova,
  B.~Mellerowicz, J.~Heldmann, C.~Stoker, et~al., Reaching 1 m deep on mars:
  the icebreaker drill, Astrobiology 13~(12) (2013) 1166--1198.

\bibitem{barbee2010comprehensive}
B.~W. Barbee, T.~Esposito, E.~Pinon~III, S.~Hur-Diaz, R.~G. Mink, D.~R. Adamo,
  A comprehensive ongoing survey of the near-earth asteroid population for
  human mission accessibility, in: Proceedings of the {AIAA}/{AAS} Guidance,
  Navigation, and Control Conference 2010, AIAA 2010-8368, 2010.

\bibitem{ethridge2016proposed}
E.~C. Ethridge, Proposed experiment for prospecting and mining water from lunar
  permafrost from boreholes using {RF} energy, in: Annual Meeting of the Lunar
  Exploration Analysis Group 2016, {LPI} Contribution No. 1960, id.5024, Vol.
  1960, 2016.

\bibitem{zacny2008drilling}
K.~Zacny, Y.~Bar-Cohen, M.~Brennan, G.~Briggs, G.~Cooper, K.~Davis, B.~Dolgin,
  D.~Glaser, B.~Glass, S.~Gorevan, et~al., Drilling systems for
  extraterrestrial subsurface exploration, Astrobiology 8~(3) (2008) 665--706.

\bibitem{Merriam2016}
E.~G. Merriam, A.~B. Berg, A.~Willig, A.~Parness, T.~Frey, L.~L. Howell,
  Microspine gripping mechanism for asteroid capture, in: Proceedings of 43rd
  Aerospace Mechanisms Symposium, 2016.

\bibitem{parness2011}
A.~Parness, Anchoring foot mechanisms for sampling and mobility in
  microgravity, in: Proceedings of 2011 IEEE ICRA, 2011.

\bibitem{bonin2016prospector}
G.~Bonin, C.~Foulds, S.~Armitage, D.~Faber, Prospector-1: The first commercial
  small spacecraft mission to an asteroid, in: Proceedings of the 30th
  {AIAA}/{USU} Conference on Small Satellites, {N}ext on the {P}ad,
  {SSC16-VI-2}, 2016.

\bibitem{Karsten2016}
J.~Karsten, M.~Bodnar, M.~Freedman, L.~Osborne, R.~Grist, R.~Hoyt, Performance
  characterization of the {HYDROS}\texttrademark water electrolysis thruster,
  in: 32nd AAS {G}uidance and {C}ontrol {C}onference, {AAS} 17-145, 2016.

\bibitem{chalex2013}
P.~Michel, M.~Barucci, A.~Cheng, H.~B{\"o}hnhardt, J.~Brucato, E.~Dotto,
  P.~Ehrenfreund, I.~Franchi, S.~Green, L.-M. Lara, et~al., Marco{P}olo-{R}:
  Near-earth asteroid sample return mission selected for the assessment study
  phase of the {ESA} program cosmic vision, Acta Astronautica 93 (2014)
  530--538.

\bibitem{wertz2011space}
J.~R. Wertz, D.~F. Everett, J.~J. Puschell, Space mission engineering: the new
  SMAD, Microcosm Press, 2011.

\bibitem{FalconH}
Spacex - capabilities and services,
  \url{http://www.spacex.com/about/capabilities}, accessed: 2018-07-02.

\bibitem{spacereport}
S.~Foundation, The Space Report: The authoritative guide to Global Space
  Activity, Space Foundation, 2017.

\end{thebibliography}

\end{document}